\documentclass[twocolumn,prd,floatfix,preprintnumbers,nofootinbib,superscriptaddress]{revtex4-2}
\usepackage{graphicx}
\usepackage{epsfig,float}
\usepackage{dcolumn}
\usepackage{bm}
\usepackage{bbm}
\usepackage{morefloats}
\usepackage{color}
\usepackage{amsmath}
\usepackage{mathrsfs}
\usepackage{slashed}
\usepackage{multirow}
\usepackage{amssymb}
\usepackage{bbding}
\usepackage{hyperref}
\hypersetup{colorlinks=true,linkcolor=blue,citecolor=blue,menucolor=black,urlcolor=black,filecolor=blue}
\usepackage{appendix}
\usepackage{multirow}
\usepackage{graphicx}
\usepackage{subfigure}
\usepackage{subfloat}   
\usepackage{float}
\usepackage{amssymb}
\usepackage{tabularx}
\graphicspath{{./image}}

\makeatletter
\newcommand*{\rom}[1]{\expandafter\@slowromancap\romannumeral #1@}
\makeatother

\usepackage{ulem}

\begin{document}

\title{Electromagnetic form factors of singly charmed baryons $\Sigma_c$ and $\Lambda_c$ in a covariant quark-diquark model}


\author{Ye Cao}~\email{caoye@impcas.ac.cn}
\affiliation{Southern Center for Nuclear-Science Theory (SCNT), Institute of Modern Physics, Chinese Academy of Sciences, Huizhou 516000, China}

\author{Cheng Chen}~\email{chencheng@impcas.ac.cn}
\affiliation{State Key Laboratory of Heavy Ion Science and Technology, Institute of Modern Physics, Chinese Academy of Sciences, Lanzhou 730000, China}
\affiliation{School of Nuclear Sciences and Technology, University of Chinese Academy of Sciences, Beijing 101408, China}

\author{Dongyan Fu}~\email{fudongyan@impcas.ac.cn}
\affiliation{Southern Center for Nuclear-Science Theory (SCNT), Institute of Modern Physics, Chinese Academy of Sciences, Huizhou 516000, China}
\affiliation{College of Physics, Guizhou University, Guiyang 550025, China}

\author{Ju-Jun Xie}~\email{xiejujun@impcas.ac.cn}
\affiliation{Southern Center for Nuclear-Science Theory (SCNT), Institute of Modern Physics, Chinese Academy of Sciences, Huizhou 516000, China}
\affiliation{State Key Laboratory of Heavy Ion Science and Technology, Institute of Modern Physics, Chinese Academy of Sciences, Lanzhou 730000, China}
\affiliation{School of Nuclear Sciences and Technology, University of Chinese Academy of Sciences, Beijing 101408, China}

\begin{abstract}

We present a systematic study of the spacelike electromagnetic form factors of the ground-state singly charmed baryons, $\Sigma_c$ ($\Sigma_c^{++},\Sigma_c^+,\Sigma_c^0$) and $\Lambda_c^+$, within a covariant quark-diquark model. Based on this framework, we obtain their magnetic moments as well as electric charge and magnetic moment radii. Such observables are important to understand the internal structure and the inner dynamics of these heavy baryon states.
Our theoretical calculations are in qualitative agreement with available lattice QCD results of $\Sigma_c^{++}$ and $\Sigma_c^0$ baryons. We also discuss the mechanism behind the dependence of the numerical results on the charm quark and light diquark. A key finding is that the electric form factors of the singly charmed baryons fall off much more slowly with momentum transfer $Q^2$ than that of the proton, indicating a more compact electric charge distribution, which is attributed to the heavy charm quark acting as a localized core. More importantly, we observe a striking difference in the magnetic structure: while the magnetic form factors of $\Sigma_c$ are dominated by the light axial-vector diquark, those of $\Lambda_c^+$ are unexpectedly governed by the charm quark due to the vanishing contribution of the scalar diquark.
This highlights the decisive role of the light-diquark spin configuration in determining the magnetic properties. Finally, using an empirical asymptotic relation connecting the spacelike and timelike regions, we predict the total cross section for $e^+e^-\to\Sigma_c\bar{\Sigma}_c$. Our findings can be tested at existing facilities including the BESIII, Belle II and LHCb, as well as the proposed Super Tau-Charm Facility.

\end{abstract}

\maketitle

\section{Introduction}

Electromagnetic form factors (EMFFs) encode the internal electromagnetic (EM) structure of hadrons~\cite{Pacetti:2014jai}. For nonpointlike particles, the electric and magnetic form factors in the spacelike region can be Fourier transformed from the momentum space to the coordinate space, thereby obtaining the distributions of electric charge and magnetic moment densities.
They also offer valuable insights into the size and shape of hadrons and some static properties like their magnetic moments.
Over the past two decades, extensive experimental and theoretical efforts have focused on the EMFFs of light mesons and baryons, especially the pion and nucleon~\cite{Hohler:1974eq, Punjabi:2015bba, Maris:2000sk, Bincer:1959tz}.
With upgraded experimental facilities, an increasing number of heavy baryons have been measured, opening new avenues for exploring their internal structure and EM properties. 


A singly heavy baryon ($Qqq$) consists of one heavy quark $Q=b, c$ and two light quarks $q=u, d, s$, with $m_Q\gg\Lambda_{\text{QCD}}\gg m_q$. This mass hierarchy allows the heavy quark to act as a static color source, with its dynamics effectively decoupled from those of the light quarks. In the limit $m_Q\to\infty$, the heavy-quark spin cannot be flipped and thus does not affect the baryon properties. This feature is known as heavy quark spin symmetry (HQSS)~\cite{Isgur:1989vq, Georgi:1990um, Isgur:1991wq}. Hence, the internal dynamics of the singly heavy baryon are governed by the light degrees of freedom and the two light quarks can be regarded as a diquark system decoupled from the heavy quark~\cite{Shifman:2024kfj}.

Under SU(3) flavor symmetry, the two light quarks in the singly heavy baryon could form either an antisymmetric flavor antitriplet $\bar{3}_F$ or a symmetric flavor sextet $6_F$. Fermi-Dirac statistics force the diquark spin to be 0 for $\bar{3}_F$ and 1 for $6_F$. Consequently, $\bar{3}_F$ baryons have total spin $J_{\bar{3}}=1/2$ (isosinglet $\Lambda_Q$, isodoublet $\Xi_{Q}$), while $6_F$ baryons can have $J_6=1/2$ or $J_6=3/2$ (isotriplet $\Sigma_Q$, isodoublet $\Xi^{\prime}_Q$, isosinglet $\Omega_Q$).
The ground-state $Qqq$ baryons are formed by coupling a heavy quark $Q$ with spin-parity $S_Q^P=1/2^+$ to a light diquark with $S_l^P=0^+$ ($\Lambda$-type) or $1^+$ ($\Sigma$-type) in a relative $S$-wave~\cite{Korner:1994nh}.
Accordingly, the wave functions are antisymmetric (under $q_1\leftrightarrow q_2$) for $\Lambda$-type,
\begin{align}
    \begin{split}
        |[q_1q_2]Q_3\rangle=|(q_1q_2)^{S_l=0}Q_3\rangle,
    \end{split}
\end{align}
and symmetric for $\Sigma$-type,
\begin{align}
    \begin{split}
        |\{q_1q_2\}Q_3\rangle=|(q_1q_2)^{S_l=1}Q_3\rangle,
    \end{split}
\end{align}
The charmed baryons $\Lambda_c^+$ and $\Sigma_c$ represent the lightest $\Lambda$-type and $\Sigma$-type heavy baryon states, respectively. Despite similar $cqq$ quark composition, they differ in the light-quark internal degrees of freedom, corresponding to scalar and axial-vector configurations. Such distinctions lead to discrepancies in their mass spectra and intrinsic structures. Hence, investigating their EMFFs is critical to unraveling the dynamics of heavy-flavor systems.

Experimentally, the spacelike and timelike EMFFs of nucleons have been extensively measured via $ep$ elastic scattering and $e^+e^-$ annihilation~\cite{Castellano:1973wh, E835:1999mlt, Antonelli:1998fv, Andivahis:1994rq, JeffersonLabHallA:1999epl, JeffersonLabHallA:2001qqe, Denig:2012by}, confirming their composite nature. In contrast, due to the short lifetime, it is difficult to produce stable and high-quality beams of $\Lambda_c^+$ and $\Sigma_c$, making spacelike EMFFs measurements very challenging. For $\Lambda_c^+$, the BESIII collaboration has measured its timelike EMFFs via the $e^+e^- \to \Lambda_c \bar{\Lambda}_c$~\cite{BESIII:2017kqg, BESIII:2023rwv, BESIII:2026qbp}, stimulating various theoretical interpretations in the timelike region~\cite{Guo:2024pti, Milstein:2022bfg, Salnikov:2023qnn, Chen:2023oqs, Chen:2024luh, Wan:2021ncg, Dai:2024lau}. Although dispersion relations provide a link between the timelike and spacelike EMFFs~\cite{Denig:2012by, Lin:2021umz}, the lack of experimental phases prevents a rigorous analytic continuation of the timelike EMFFs to the spacelike region without loss of physical information. As for $\Sigma_c$, no direct experimental measurements of their EMFFs exist in either region.

Thus, a detailed examination of the spacelike EMFFs of $\Lambda_c^+$ and $\Sigma_c$ is crucial for understanding heavy-flavor dynamics and distinguishing the EM features between light and heavy baryons. However, such efforts are currently hindered by a total absence of experimental data and a scarcity of theoretical insights. Lattice QCD (LQCD) calculations for $\Sigma_c^{++}$ and $\Sigma_c^0$ have been performed by Can et al.~\cite{Can:2013tna, Can:2021ehb}, while Kim et al.~\cite{Kim:2018nqf, Kim:2019wbg, Kim:2025bms} estimated the EMFFs of both $\Sigma_c$ and $\Lambda_c^+$ using the chiral quark-soliton model, where the form factor of $\Sigma^{++}_c$ is compared with that from LQCD. In addition to the studies of the spacelike EMFFs of $\Sigma_c$ and $\Lambda_c^+$ in Refs.~\cite{Can:2013tna, Can:2021ehb, Kim:2018nqf, Kim:2019wbg, Kim:2025bms, Liu:2018tqe}, the magnetic moment, being one of the most interesting observables, has also been investigated. Under the usual assumption that the baryon magnetic moments arise purely from their constituent quarks~\cite{Franklin:1981rc}, one has
\begin{equation}
\mu_B=\sum_i\langle B\uparrow|\mu_i \hat{\sigma}_i^z|B\uparrow\rangle ,
\end{equation}
where $|B\uparrow\rangle$ is the SU(6) spin-flavor baryon wave function and $\mu_i=\frac{e_i}{2m_i}$~\footnote{The index $i=c,u,d$ labels the constituent quarks inside the baryon.} with $e_i$, $\hat{\sigma}_i$ and $m_i$ being the electric charge, Pauli operator and constituent mass of the $i$-th quark in the baryon. This formula yields
\begin{align}
    \begin{split}
        \mu_{\Sigma_c^{++}}&=\frac13(4\mu_u-\mu_c),\ \ \ \mu_{\Sigma_c^{+}}=\frac13(2\mu_u+2\mu_d-\mu_c),\\
        \mu_{\Sigma_c^{0}}&=\frac13(4\mu_d-\mu_c),\ \ \ \ \mu_{\Lambda_c^+}=\mu_c.
        \label{eq:magnetic moment}
    \end{split}
\end{align}
Since $\mu_i\propto 1/m_i$, the magnetic moment of the singly charmed baryon is generally governed by light quarks in the limit $m_c\to\infty$. Various theoretical approaches have been employed to compute these quantities, including the effective mass scheme (EMS)~\cite{Mohan:2022sxm}, chiral perturbation theory ($\chi$PT)~\cite{Shi:2018rhk, Wang:2018gpl}, bag model (BM)~\cite{Bernotas:2013eia, Sharma:2010vv}, lattice QCD (LQCD)~\cite{Can:2013tna, Can:2021ehb}, light cone QCD sum rule (LCQSR)~\cite{Ozdem:2024brk, Aliev:2015axa, Aliev:2001ig}, relativistic quark model (RQM)~\cite{Julia-Diaz:2004yqv, Faessler:2006ft},  potential model (PM)~\cite{Barik:1983ics}, hypercentral constituent quark model (hCQM)~\cite{Patel:2025txc, Gandhi:2018lez}, quark-diquark model (QDM)~\cite{Farhadi:2023ucs}, chiral quark-soliton model ($\chi$QSM)~\cite{Kim:2018nqf, Yang:2018uoj}, and Skyrme model (SM)~\cite{Scholl:2003ip, Oh:1991ws}. Given the absence of experimental data, it is particularly instructive to compare predictions from these different approaches.

In this paper, we study the EMFFs of the singly charmed baryons $\Sigma_c$ ($\Sigma_c^{++},\Sigma_c^+,\Sigma_c^0$) and $\Lambda_c^+$ in a covariant quark-diquark model. The electric charge and magnetic moment radii, and the magnetic moments of these baryons are also investigated.
The remainder of this paper is organized as follows. In Sec.~\ref{sec-Formalism}, we introduce the EMFFs of spin-1/2 baryons and the covariant quark–diquark model.
Numerical results for the spacelike EMFFs and extracted EM properties are presented in Sec.~\ref{sec-result}.
Finally, a brief summary is given in Sec.~\ref{sec-summary}.

\section{Formalism}
\label{sec-Formalism}

\subsection{Electromagnetic form factors}

In the one photon approximation, the EM current matrix element for the spin-1/2 baryon is given by
\begin{align}
\begin{split}
    &\langle p^{\prime},\lambda^{\prime}|\hat{J}^{\mu}(0)|p,\lambda\rangle = \\
    & \bar{u}({p}^{\prime},\lambda^{\prime})\Bigg[F_1(t)\gamma^{\mu}+F_2(t)\frac{i\sigma^{\mu\nu}\Delta_{\nu}}{2M}\Bigg]u(p,\lambda),
    \end{split}
\end{align}
where $p(p^{\prime})$ and $\lambda(\lambda^{\prime})$ are the momentum and helicity of the initial (final) baryon, $t=\Delta^2=(p^{\prime}-p)^2<0$ is the squared transfer momentum in the spacelike region, $M$ is the baryon mass, and $u(p,\lambda)$ is the Dirac spinor with normalization as $\bar{u}(p,\lambda^{\prime}) u(p,\lambda) = 2M\delta_{\lambda^{\prime}\lambda}$. The Sachs form factors $G_E(t)$ and $G_M(t)$ are related to the Dirac and Pauli form factors $F_1(t)$ and $F_2(t)$~\cite{Ernst:1960zza},
\begin{align}
    \begin{split}
        G_E(t) &=F_1(t)+\tau F_2(t), \\ 
        G_M(t) &=F_1(t)+F_2(t),
    \end{split}
\end{align}
with $\tau={t}/({4M^2})$. In the forward limit $t \to 0$, one obtains the charge $\mathcal{Q}=G_E(0)$ and the magnetic moment $\mu=\frac{e}{2M}G_M(0) = \frac{M_N}{M} G_M(0) \mu_N$ (in the unit of nuclear magneton $\mu_N$ and $M_N$ is the nucleon mass). {Moreover, the slope of an EMFF at $t=0$ provides a measure of either the mean square electric charge radius or the magnetic moment radius of a hadron. Following the convention of Ref.~\cite{Cloet:2014rja}, all mean square radii can be defined by the unified formula,
\begin{align}
\langle r^2 \rangle = \frac{6}{\eta} \left. \frac{\partial f(t)}{\partial t} \right|_{t=0}, \quad \eta = 
\begin{cases} 
1 & \text{if } f(0) = 0, \\ 
f(0) & \text{if } f(0) \neq 0, 
\end{cases}
\label{eq:radii_general}
\end{align}
where $f(Q^2)$ is an arbitrary form factor. This unified prescription naturally accommodates both charged and neutral particles on the same footing. For a charged particle, Eq.~\eqref{eq:radii_general} reduces to the common forms, 
\begin{align}
    \begin{split}
        \langle r_E^2\rangle &= \frac{6}{G_E(0)}\frac{d}{dt}G_E(t)\Big|_{t=0}, \\
        \langle r_M^2\rangle &= \frac{6}{G_{M}(0)}\frac{d}{dt}G_M(t)\Big|_{t=0}.
        \label{eq:radii}
    \end{split}
\end{align}
However, for the neutral particle, the definition of the electric charge radius becomes $\langle r_E^2\rangle=6\frac{d}{dt}G_E(t)\big|_{t=0}$, where the normalization factor $G_E(0)$ is removed from the denominator to avoid the divergence. This convention is widely adopted in the literature for neutral hadrons~\cite{Julia-Diaz:2004yqv, Cloet:2014rja, Kim:2018nqf, Fu:2022rkn, Wang:2023bjp}. 

}

\subsection{Quark-diquark model}


{
The diquark model is often employed to simplify systems composed of more than two particles, since the reduced degrees of freedom make the analysis and numerical treatment more tractable~\cite{Barabanov:2020jvn} at a practical level~\cite{Barabanov:2020jvn}. In the quark-diquark picture, a baryon is described as a constituent diquark plus a quark. Over the past decades, this picture has proven useful in calculating baryon spectra, including both light baryons~\cite{Santopinto:2004hw,Galata:2012xt, Lichtenberg:1982jp, Ferretti:2011zz, Santopinto:2014opa,Gutierrez:2014qpa, DeSanctis:2014ria} and heavy-light baryons~\cite{Ebert:1996ec, Gershtein:2000nx, Ebert:2002ig, Giannuzzi:2009gh,Ali:2017wsf, Santopinto:2018ljf}, as well as in studies of the baryon structure, such as the EMFFs of light octet and decuplet baryons~\cite{Santopinto:2004hw, Jakob:1993th, Keiner:1995bu, Ma:2002ir,DeSanctis:2011zz, Fu:2022rkn, Fu:2023ijy, Wang:2023bjp, Wang:2024abv,Fu:2025vkq} and EM transition form factors~\cite{Santopinto:2004hw, Ramalho:2008ra, Wang:2026nlr}. Given these successful applications, it is natural to extend this framework to the EMFFs of heavy baryons, especially $Qqq$ systems. Therefore, the heavy quark can be treated as a pointlike object with the definite color, spin and mass, while the two light quarks are strongly correlated into a diquark due to HQSS~\cite{Isgur:1989vq, Georgi:1990um, Isgur:1991wq}. In this work, we employ the quark-diquark model to investigate the EM structures of the lightest charmed ground states, $\Sigma_c$ and $\Lambda_c^+$, which are composed of a charm quark and a light diquark (a pair of $uu$, $ud$, or $dd$) in a relative $S$-wave. This framework also separates the heavy and light degrees of freedom in a physically transparent way, allowing us to examine their respective roles in shaping the EM structures and properties of the $\Sigma_c$ and $\Lambda_c^+$.

}

\begin{figure}[htbp]
    \centering
    \includegraphics[scale=0.55]{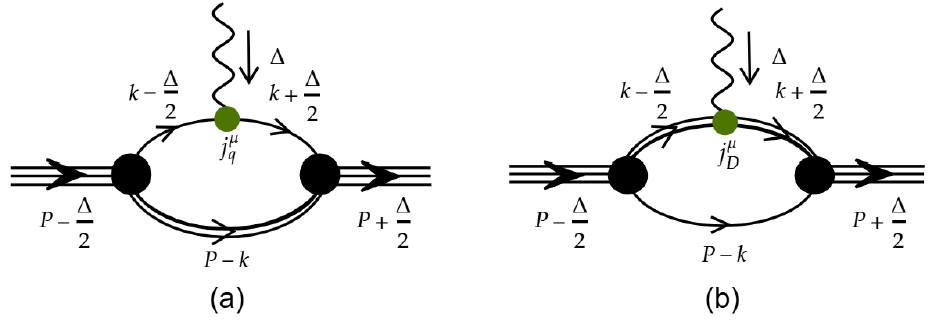}
    \caption{Feynman diagrams for photon coupling to a quark (a) and diquark (b). In (b), the photon interacts with each quark inside the nonpointlike diquark.} \label{fig:matrix element of EM current}
\end{figure}

As shown in Fig.~\ref{fig:matrix element of EM current}, the matrix element of the EM current is the sum of the quark and diquark contributions,
\begin{align}
    \begin{split}
        \langle p^{\prime},\lambda^{\prime}|\hat{J}^{\mu}|p,\lambda\rangle=\langle p^{\prime},\lambda^{\prime}|\hat{J}^{\mu}_q|p,\lambda\rangle+\langle p^{\prime},\lambda^{\prime}|\hat{J}^{\mu}_D|p,\lambda\rangle,
    \end{split}
\end{align}
where the subscripts $q$ and $D$ refer to quark (not light quark) and diquark, respectively. To analyze the EM structure, we work in the Breit frame, in which the momenta of the initial and final baryons are defined as $p=(E, -{\boldsymbol{\Delta}}/{2})$ and $p^{\prime}=(E,{\boldsymbol{\Delta}}/{2})$, respectively. The average four-momentum and the four-momentum transfer are then given by $P=(p+p^{\prime})/{2}=(E,\boldsymbol{0})$ and $\Delta= p' - p = (0,\boldsymbol{\Delta})$. Then, the quark contributions for the EMFFs of $\Sigma_c$ and $\Lambda_c^+$ can be written as
\begin{align}
    \begin{split}
        &\langle p^{\prime},\lambda^{\prime}|\hat{J}_q^{\mu}(0)|p,\lambda\rangle_{\Sigma_c} = i \mathcal{Q}_qe\bar{u}(p^{\prime},\lambda^{\prime})\int\frac{d^4k}{(2\pi)^4 \mathcal{D}}\\
        &\times \tilde{\Gamma}_a^{\beta^{\prime}}\Big(\slashed{k}+\frac{\slashed{\Delta}}{2}+m_q\Big)\gamma^{\mu}\Big(\slashed{k}-\frac{\slashed{\Delta}}{2}+m_q\Big)\\
        &\times\Bigg[g_{\beta^{\prime}\beta} - \frac{(P_{\beta^{\prime}}-k_{\beta^{\prime}})(P_{\beta}-k_{\beta})}{m_D^2}\Bigg]\tilde{\Gamma}_a^{\beta}u(p,\lambda),\\
        &\langle p^{\prime},\lambda^{\prime}|\hat{J}_q^{\mu}(0)|p,\lambda\rangle_{\Lambda_c^+} = -i \mathcal{Q}_qe\bar{u}(p^{\prime},\lambda^{\prime}) \int\frac{d^4k}{(2\pi)^4 \mathcal{D}}\\
        &\times \tilde{\Gamma}_s\Big(\slashed{k}+\frac{\slashed{\Delta}}{2}+m_q\Big)\gamma^{\mu}\Big(\slashed{k}-\frac{\slashed{\Delta}}{2}+m_q\Big)\tilde{\Gamma}_su(p,\lambda),
        \label{eq:quark contribution}
    \end{split}
\end{align}
where $\mathcal{Q}_q$ is the charge number of the interacting quark, and $\mathcal{D}$ is the product of all propagator denominators,
\begin{eqnarray}
\mathcal{D} &=& \bigg[\Big(k+\frac{\Delta}{2}\Big)^2-m_q^2+i\epsilon\bigg]\bigg[\Big(k-\frac{\Delta}{2}\Big)^2-m_q^2+i\epsilon\bigg] \nonumber \\
 && \times\bigg[\Big(k-P\Big)^2-m_{D}^2+i\epsilon\bigg], \label{eq:propagator}
\end{eqnarray}
with $m_D$ the diquark mass. The baryon-quark-diquark vertices in Eq.~(\ref{eq:quark contribution}) are $\tilde{\Gamma}_a^{\mu}=\Gamma_a^{\mu}\Xi$ and $\tilde{\Gamma}_s=\Gamma_s\Xi$, with subscript $a(s)$ for the axial-vector (scalar) diquark. Following Ref.~\cite{Scadron:1968zz}, the Lorentz structures of $\Gamma_a^{\mu}$ and $\Gamma_s$ are
\begin{align}
    \Gamma_a^{\mu}=c_a(\gamma^{\mu}+g_1\frac{p_r^{\mu}}{M})\gamma^5,\ \ \ \Gamma_s=c_s,
    \label{eq:g1}
\end{align}
where $p_r$ is the quark-diquark relative momentum. {The vertex structures are dictated by the covariance, angular momentum conservation and parity conservation.} The parameter $g_1$ is a coupling constant to be determined by fitting to the experimental data or the LQCD calculations. An additional scalar function $\Xi(p_1,p_2)$ is introduced at the baryon-quark-diquark interaction vertex for the regularization of the loop integral~\cite{Frederico:2009fk},
\begin{align}
    \begin{split}
        \Xi(p_1,p_2)=\frac{c}{(p_1^2-m_R^2+i\epsilon)(p_2^2-m_R^2+i\epsilon)},
        \label{eq:scalar function}
    \end{split}
\end{align}
with $p_1$ ($p_2$) the quark (diquark) momentum, and $m_R$ a cutoff parameter. {As discussed in Refs.~\cite{Davidson:1994uv, Broniowski:2008hx}, regularization via a momentum-dependent scalar function breaks the gauge invariance and violates the EM Ward-Takahashi identities. Despite this formal deficiency, we adopt this cutoff scheme to suppress high-momentum
contributions and render the loop integrals finite. It has since been employed in a wide
range of calculations for mesons and almost all light baryons~\cite{Frederico:2009fk, Sun:2017gtz, Fu:2022rkn, Wang:2023bjp,
Wang:2024abv, Fu:2025upc, Wang:2026nlr}, with predictions in good agreement with experimental data and other theoretical approaches. These studies
demonstrate that, despite the explicit breaking of gauge invariance, a momentum-dependent scalar regulator such as Eq.~(\ref{eq:scalar function}) provides a simple and effective means of describing hadron structure and properties in the low-energy region.}

\begin{figure}[htbp]
    \centering
    \includegraphics[scale=0.5]{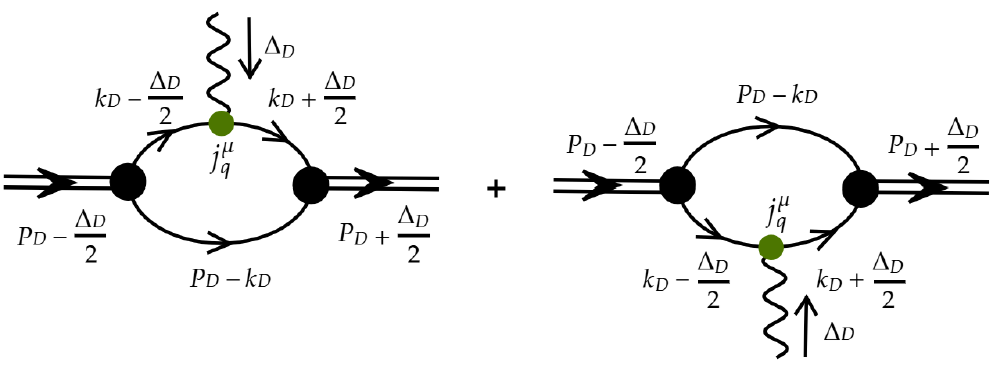}
    \caption{The effective EM coupling of a photon to a diquark is given by the sum of the photon interactions with each quark inside the diquark.} \label{fig:diquark EM current}
\end{figure}

Next, the diquark contributions can be obtained, which are
\begin{align}
    \begin{split}
        &\langle p^{\prime},\lambda^{\prime}|\hat{J}_{D}^{\mu}(0)|p,\lambda\rangle_{\Sigma_c}= -i \mathcal{Q}_De\bar{u}(p^{\prime},\lambda^{\prime}) \\
        &\times \int\frac{d^4k}{(2\pi)^4 \mathcal{D}^{\prime}} \tilde{\Gamma}_a^{\beta^{\prime}}\Big(\slashed{P}-\slashed{k}+m_q\Big)\tilde{\Gamma}_a^{\beta}\\
        &\times\Bigg[g_{\alpha^{\prime}\beta^{\prime}}- \frac{(k_{\alpha^{\prime}}+\frac{\Delta_{\alpha^{\prime}}}{2})(k_{\beta^{\prime}}+\frac{\Delta_{\beta^{\prime}}}{2})}{m_D^2}\Bigg]j_{D(a)}^{\mu,\alpha^{\prime}\alpha}\\
        &\times\Bigg[g_{\alpha\beta} - \frac{(k_{\alpha}-\frac{\Delta_{\alpha}}{2})(k_{\beta}-\frac{\Delta_{\beta}}{2})}{m_D^2}\Bigg]u(p,\lambda),\\
        &\langle p^{\prime},\lambda^{\prime}|\hat{J}_{D}^{\mu}(0)|p,\lambda\rangle_{\Lambda_c^+} = -i\mathcal{Q}_De\bar{u}(p^{\prime},\lambda^{\prime})\\
        &\times\int\frac{d^4k}{(2\pi)^4 \mathcal{D}^{\prime}} \tilde{\Gamma}_s\Big(\slashed{P}-\slashed{k}+m_q\Big)j_{D(s)}^{\mu}\tilde{\Gamma}_su(p,\lambda), \label{eq:diquark contribution}
    \end{split}
\end{align}
where $\mathcal{Q}_D$ is the diquark charge, and $\mathcal{D}^{\prime}$ is obtained from $\mathcal{D}$  by exchanging $m_q$ and $m_D$. Here, $j_{D(a)}^{\mu,\alpha^{\prime}\alpha}$ and $j_{D(s)}^{\mu}$ are the effective EM currents for the axial-vector diquark and the scalar diquark, respectively. These currents arise from the diquark's internal structure, as illustrated in Fig.~\ref{fig:diquark EM current}, can be obtained as,
\begin{align}
    \begin{split}
    & -\epsilon_{\alpha^{\prime}}^*(p_D^{\prime},\lambda_D^{\prime})j_{D(a)}^{\mu,\alpha^{\prime}\alpha}\epsilon_{\alpha}(p_D,\lambda_D) \\
    & = \sum_q\langle p_{D}^{\prime},\lambda_D^{\prime}|\hat{J}_q^{\mu}|p_D,\lambda_D\rangle,\\ 
    & j_{D(s)}^{\mu} = \sum_q\langle p_{D}^{\prime},\lambda_D^{\prime}|\hat{J}_q^{\mu}|p_D,\lambda_D\rangle,
   \label{eq:j_D}
    \end{split}
\end{align}
where $\epsilon_{\alpha}(p_D,\lambda_D)$ is the spin-1 diquark polarization vector, and $p_D$ ($p_D^{\prime}$) and $\lambda_D $ $(\lambda'_D)$ denote the momentum and the helicity of the initial (final) diquark. We also introduce the kinematic variables: $P_D^{\mu}=(p_D^{\mu}+p_D^{\prime\mu})/2$, $\Delta_D^{\mu}=p_D^{\prime\mu}-p_D^{\mu} = \Delta^\mu$, and $\Delta_D^2=t_D=t$. 

The effective Lagrangians for the quark-quark-diquark vertices are taken from Ref.~\cite{Meyer:1994cn},
\begin{align}
    \mathcal{L}_{D(a)\to qq}&=c_{D(a)}\psi_q^TC^{-1}\gamma^{\mu}\psi_q A_{D \mu}\Xi_D+h.c.,\\
    \mathcal{L}_{D(s)\to qq}&=c_{D(s)}\psi_q^TC^{-1}\gamma_{5}\psi_q \phi_D \Xi_D+h.c.,
\end{align}
where $\psi_q^T$ represents the charge-conjugate of the quark field, $A_{D \mu}$ ($\phi_D$) denotes the axial-vector (scalar) diquark field, and $C$ is the charge-conjugate oprator $C=i\gamma^2\gamma^0$. Here, we take $\Xi_D$ as the same form as $\Xi$ with the same cutoff $m_R$ to reduce the parameter.

The quark loop contribution for the scalar diquark is
\begin{align}
    \begin{split}
        &\langle p_{D}^{\prime},\lambda_D^{\prime}|\hat{J}_q^{\mu}(0)|p_D,\lambda_D\rangle_s=\mathcal{Q}_qe\int\frac{d^4k_D}{(2\pi)^4}\frac{-ic^2c_{D(s)}^2}{\mathscr{D}_D} \\ 
        & \times \text{Tr}\Bigg[{\gamma^5}
        \Big(\slashed{k}_D+\frac{\slashed{\Delta}}{2}+m_q\Big)\gamma^{\mu}\Big(\slashed{k}_D-\frac{\slashed{\Delta}}{2}+m_q\Big){\gamma^5} \\
        & \times \Big({\slashed{k}_D-\slashed{P}_D}+m_q\Big)\Bigg],
        \label{eq:diquark’s internal structure loop}
    \end{split}
\end{align}
where $\mathscr{D}_D$ has the similar function as $\mathcal{D}$.
The axial-vector contribution is derived analogously by substituting $\gamma^5$ with $\gamma^{\alpha}$ and performing the index contraction. 

Finally, the diquark currents in Eq.~(\ref{eq:j_D}) are parametrized as
\begin{align}
\begin{split}
    j_{D(a)}^{\mu,\alpha^{\prime}\alpha} =&  2 \Big[g^{\alpha^{\prime}\alpha}F_{D;1}^V(t)-\frac{\Delta^{\alpha^{\prime}}\Delta^{\alpha}}{2m_D^2}F_{D;3}^V(t)\Big] P_D^{\mu} \\
   & - (\Delta^{\alpha^{\prime}}g^{\mu\alpha}-\Delta^{\alpha}g^{\mu\alpha^{\prime}})F_{D;2}^V(t),\\
    j_{D(s)}^{\mu}= & 2 f_D(t) P_D^{\mu},
    \end{split}
\end{align}
where $F_{D;1,2,3}^V(t)$ and $f_D(t)$ are the EMFFs of the spin-1 and spin-0 particles, respectively. The combined products $cc_{D(a)}$ and $cc_{D(s)}$ are fixed by charge normalization of diquarks and are not free parameters.

\section{Numerical results and discussions} \label{sec-result}

\subsection{Model parameters}
\label{sec:parameters}

Before performing numerical calculations, it is necessary to fix the model parameters. There are three types, the constituent masses ($m_c$ for charm quark, $m_q$ with $q=u,d$ for up and down quarks, and $m_{D}$ with $D=[qq]$ or $\{qq\}$ for light diquark), the coupling $g_1$ in the baryon-quark-diquark vertex [see Eq.~(\ref{eq:g1})], and the cutoff parameter $m_R$.

For the charmed baryon mass $M$~\footnote{To ensure a bound state of the quark and diquark, the input masses must satisfy $M<m_c+m_D$ and $m_D<2m_q$.}, we adopt the physical values $M_{\Sigma_c}=2.453$ GeV and $M_{\Lambda_c^+}=2.286$ GeV. {In the constituent quark-diquark model, the quark and diquark masses are effective parameters that absorb nonperturbative QCD effects and are typically determined by calculations of the baryon spectra.
Their values may vary across different phenomenological implementations. Since our primary goal is not to reproduce the mass spectrum of charmed baryons, we choose the constituent quark masses within ranges commonly adopted by previous works~\cite{Capstick:1986ter,Lichtenberg:1996fi,Ebert:1996ab,Ebert:2005xj,Ebert:2011kk,Chen:2016iyi,Farhadi:2023ucs,Ni:2021pce}, where $m_c$ and $m_q$ are typically 1.5--1.7~GeV and 300--500~MeV, respectively. Within these ranges, we fine-tune $m_c$ and $m_q$ to provide the best qualitative description of the LQCD results for $\Sigma_c^{++}$ and $\Sigma_c^0$ reported in Refs.~\cite{Can:2013tna, Can:2021ehb}.

\begin{table}[htbp]
    \centering
    \caption{Mass difference (in MeV) between scalar ($[qq]$) and axial-vector diquark ($\{qq\}$) with $q=u,d$.}
    \begin{tabular}{ccc|ccc}
    \hline\hline
         $m_{[qq]}$  &$m_{\{qq\}}-m_{[qq]}$  &Ref. &
         $m_{[qq]}$  &$m_{\{qq\}}-m_{[qq]}$  &Ref. \\
    \hline
         - &210 &\cite{Jaffe:2004ph} &
         691 &149 &\cite{Ferretti:2019zyh} \\
         595 &205 &\cite{Lichtenberg:1996fi} &
         740 &210 &\cite{Hecht:2002ej} \\
         688 &202 &\cite{Maris:2002yu} &
         710 &199 &\cite{Ebert:2005xj} \\
         600 &350 &\cite{Ferretti:2011zz} &
         737 &212 &\cite{Burden:1996nh} \\
         696 &110 &\cite{Hess:1998sd} &
         705 &190 &\cite{Ebert:1996ab} \\
    \hline\hline
    \end{tabular}
    \label{tab:mass difference}
\end{table}

The diquark mass is less than the sum of two quarks and the axial-vector diquark mass is heavier than the scalar diquark mass, i.e., $m_{\{qq\}} > m_{[qq]}$, due to the different spin-spin interaction between the two quarks. In the $\Lambda_c^+$ baryon, the two light quarks form a scalar diquark ($[qq]$), while in the $\Sigma_c$ baryons they form an axial-vector diquark ($\{qq\}$). According to HQSS, the charm quark couples weakly to the light quarks in a $cqq$ system, so the light-diquark mass needs to be adjusted when moving from $\Sigma_c$ to $\Lambda_c^+$. We adopt the following strategy: we first set the mass splitting $m_{\{qq\}}-m_{[qq]}=M_{\Sigma_c}-M_{\Lambda_c^+}\approx 170$~MeV, and then vary this difference in the range 160--180 MeV to estimate the uncertainty introduced by this parameter. For comparison, we list the diquark mass splittings obtained from various theoretical approaches~\cite{Jaffe:2004ph, Ferretti:2019zyh, Lichtenberg:1996fi, Hecht:2002ej, Maris:2002yu, Roberts:2011cf, Ferretti:2011zz, Burden:1996nh, Hess:1998sd, Santopinto:2014opa} in Tab.~\ref{tab:mass difference}.
The range of 100--350~MeV reflects the model dependence of the parameters and our chosen range of 160--180~MeV is well within this spread.
}

The remaining parameters for the $\Sigma_c$ system are fitted to LQCD data as $g_1 = 0.7$ and $m_R =3.1$ GeV. Our $g_1$ value agrees closely with Ref.~\cite{Wang:2024abv}, and the selected $m_R$ also satisfies the convention of being larger than the baryon mass, as supported by Refs.~\cite{Fu:2022rkn,Fu:2023ijy,Wang:2023bjp,Wang:2024abv}. The same $m_R$ is used for $\Lambda_c^+$ due to the similar masses of $\Lambda_c^+$ and $\Sigma_c$. {We have also systematically verified the numerical stability of our
results against variations of the cutoff parameter $m_R$ within a reasonable range (2.8--3.4~GeV) in the following subsection, and find that the extracted observables are not sensitive to the specific choice of $m_R$.}



\begin{table}[htpb]
    \centering
    \caption{Parameters used in this work.}
    \begin{tabular}{c c c c c c c}
    \hline\hline
    $m_c$ (GeV)  &$m_q$ (MeV) &$m_{\{qq\}}$ (MeV) &$m_{[qq]}$ (MeV) \\
    \hline                  $1.62$ &$450$ &$860$ &$690$ \\
        \hline\hline
    \end{tabular}
    \label{tab:mass para}
\end{table}



{We assign physically acceptable values of the masses to the different quarks and diquarks, as listed in Tab.~\ref{tab:mass para}. We also vary these input masses with reasonable ranges while preserving the bound-state conditions $M<m_c+m_D$ and $m_D<2m_q$. The results of this sensitivity analysis are presented in the following subsection.} Furthermore, the coefficients $cc_{a/s}$ and $cc_{D(a/s)}$ can be determined by imposing the normalization condition of the electric charge, which gives
\begin{eqnarray}
\big(cc_{a},\ cc_{s}\big) &=& (97.0,\ 177.9)\ \text{GeV}^4,\\
\big(cc_{D(a)},cc_{D(s)}\big) &=&  (349.4,\ 486.8)\ \text{GeV}^4.
\end{eqnarray}
Notably, the obtained values are larger than those for nucleons reported in Ref.~\cite{Wang:2024abv}. This discrepancy is primarily attributed to the inverse tenth power of the baryon mass emerging from the integration of Eqs.~(\ref{eq:quark contribution}) and (\ref{eq:diquark contribution}).

\subsection{EMFFs of $\Sigma_c$ and $\Lambda_c$ in the spacelike region}

\begin{figure*}[htbp]
    \centering
    \includegraphics[scale=0.42]{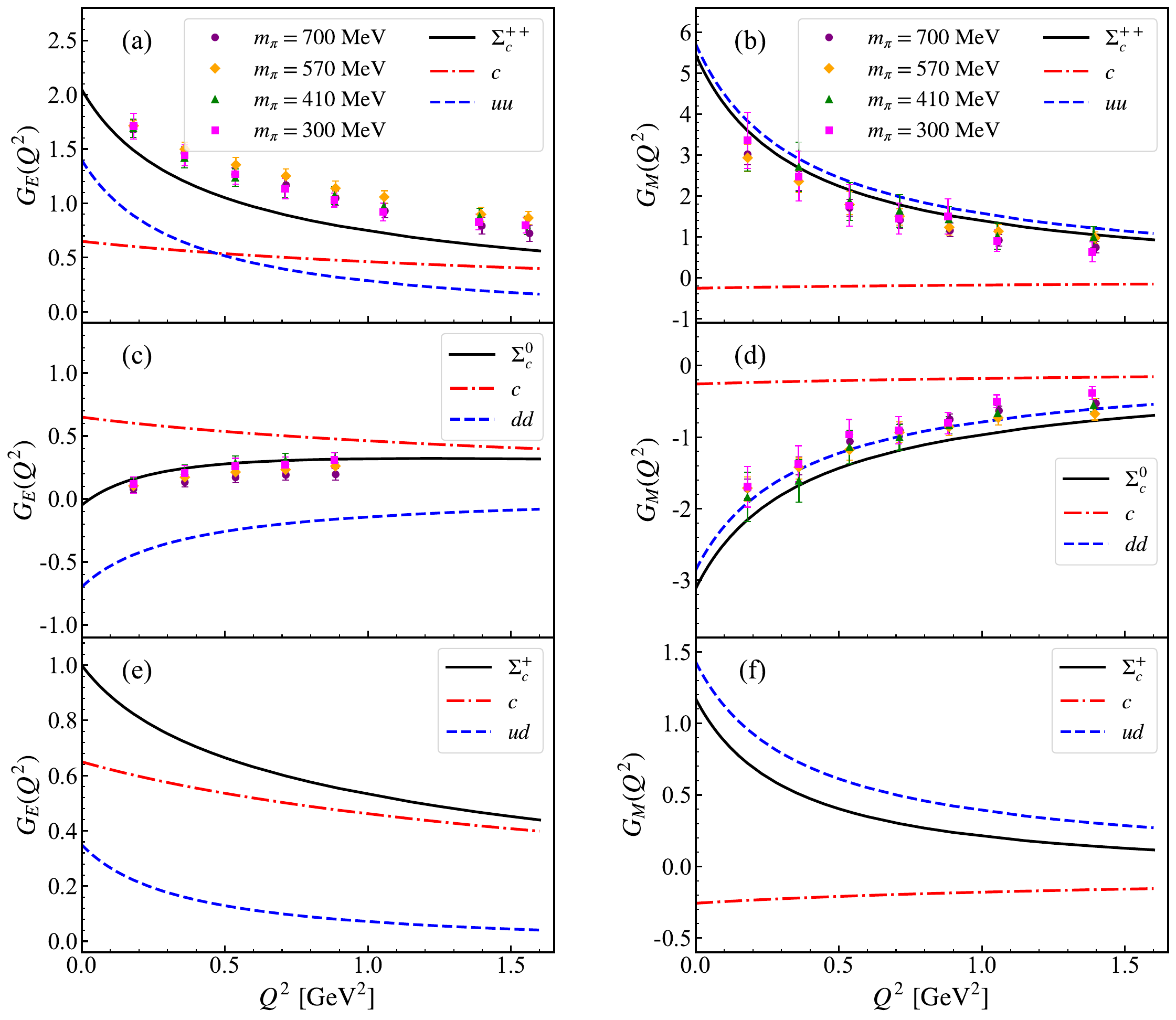}
    \caption{Electric form factors $G_E (Q^2)$ (left panel) and magnetic form factors $G_M (Q^2)$ (right panel) of $\Sigma_c^{++}$, $\Sigma_c^0$ and $\Sigma_c^+$ as functions of the squared momentum transfer $Q^2=-t$, compared with LQCD results~\cite{Can:2013tna,Can:2021ehb}. The black solid, red dash-dotted, and blue dashed curves correspond to the total, charm quark, and light diquark contributions, respectively.}
    \label{fig:Sigma_c EMFFs}
\end{figure*}

Following the on-shell identities of Ref.~\cite{Cotogno:2019vjb}, we extract the EMFFs from the integrals in Eqs.~(\ref{eq:quark contribution}) and (\ref{eq:diquark contribution}). Our theoretical results of the EMFFs are illustrated in Fig.~\ref{fig:Sigma_c EMFFs}, where the numerical results for $\Sigma_c^{++}$ and $\Sigma_c^0$ are compared with the LQCD data, while the results for $\Sigma_c^+$ are obtained using the same set of model parameters. In Fig.~\ref{fig:Sigma_c EMFFs}, the dots represent the LQCD results at various nonphysical pion masses, which are close to each other, while our black solid curves at physical pion mass are consistent with the LQCD data. Our numerical results for the electric form factor $G_E(Q^2)$ ($Q^2=-t$ is the squared momentum transfer) of $\Sigma^{++}_c$ decrease slightly more rapidly than the LQCD calculations. Nevertheless, our calculations are consistent with the finding presented in Ref.~\cite{Kim:2018nqf}. This behavior is in agreement with a well-established feature observed for the nucleon electric form factors: experimental measurements in Refs.~\cite{Castellano:1973wh, E835:1999mlt, Antonelli:1998fv, Andivahis:1994rq,JeffersonLabHallA:1999epl, JeffersonLabHallA:2001qqe, Denig:2012by} also fall off faster than those obtained by the LQCD calculations at unphysical pion masses~\cite{Capitani:2015sba,Abdel-Rehim:2015jna,Djukanovic:2015hnh,Alexandrou:2017ypw}.

It is worth noting that the $G_E(0)$ of $\Sigma_c^+$ is normalized to its charge, which is one in unit of $e$, and the same normalization coefficient is applied to $\Sigma_c^{++}$ and $\Sigma_c^0$. As shown in Fig.~\ref{fig:Sigma_c EMFFs}, the resulting $G_E(0)$ for $\Sigma_c^{++}$ and $\Sigma_c^0$ are not exactly two and zero (in unit of $e$), but are extremely close to these two values. This demonstrates that one can get an unified description of differently charged baryons within the same multiplet under the same theoretical model parameters. It is also found that the electric form factor $G_E$ of the singly positive-charged charmed baryons $\Sigma_c^+$ (and similarly for $\Lambda_c^+$) falls off much more slowly with $Q^2$ than that of the proton. This reveals a profound physical picture: charmed baryons are electrically compact objects, implying that their intrinsic size is significantly smaller than that of the proton. This feature will become more evident when we discuss the electric radii later.

Since the EM current is decomposed into heavy and light parts, the heavy-quark contribution and light-diquark contribution can be calculated separately within the quark-diquark picture. Their respective contributions to the EM properties of $\Sigma_c$ are investigated to gain deeper insight into the underlying quark dynamics. Their individual contributions are also presented in Fig.~\ref{fig:Sigma_c EMFFs}. Taking $\Sigma_c^{++}$ as an example, the ratio of quark to diquark contributions to $G_E$ approaches $\mathcal{Q}_c/\mathcal{Q}_{uu}=1/2$ as $Q^2\to 0$. This approximate relation also holds for $G_E$ of $\Sigma_c^+$ ($\mathcal{Q}_c/\mathcal{Q}_{ud}=2$) and $\Sigma_c^0$ ($\mathcal{Q}_c/\mathcal{Q}_{dd}=-1$) as shown in the left panel of Fig.~\ref{fig:Sigma_c EMFFs}. This indicates that when the $Q^2$ is small, the EM interaction probes the
diquark as an effectively pointlike particle, which is consistent with the physical intuition. However, as $Q^2$ increases, the EM current begins to resolve the internal structure of the diquark, making the effects of its binding and quark substructure increasingly significant. We also find that the charm quark's contribution to $G_E$ decreases slowly with increasing $Q^2$, consistent with the approximation adopted in Ref.~\cite{Kim:2018nqf}, where it is treated as a constant under the assumption of an infinitely heavy charm quark. However, our framework takes into account the finite constituent mass of the charm quark and thus reveals a mild $Q^2$-dependence of this contribution.

We now turn to the magnetic form factor $G_M$ of $\Sigma_c$. Since the magnetic moment of the heavy quark is proportional to the inverse of its mass, $\mu_c\sim (Q_c/m_c)\sigma$, $G_M$ of the singly heavy baryons are predominantly governed by the light quarks. As shown in the right panel of Fig.~\ref{fig:Sigma_c EMFFs}, the contribution from the light diquark to $G_M$ of $\Sigma_c$ is substantially larger than that from the charm quark. This feature indicates that the spatial distribution of light quarks extends significantly farther than that of the charm quark. In other words, the charm quark acts as a heavy core. It is also observed from Fig.~\ref{fig:Sigma_c EMFFs} that the contributions of the charm quark and the light diquark to $G_M$ (or, equivalently, their magnetic moments at $Q^2=0$) have opposite signs. After removing the respective electric charge factors, it becomes clear that the spins of the heavy and light components in all three $\Sigma_c$ baryons are oppositely oriented, implying that they are predominantly antialigned to each other. This characteristic can also be directly observed from Eq.~(\ref{eq:magnetic moment}) and is in agreement with the findings of lattice QCD~\cite{Can:2013tna,Can:2021ehb}.

\begin{figure}[htbp]
    \centering
    \includegraphics[scale=0.42]{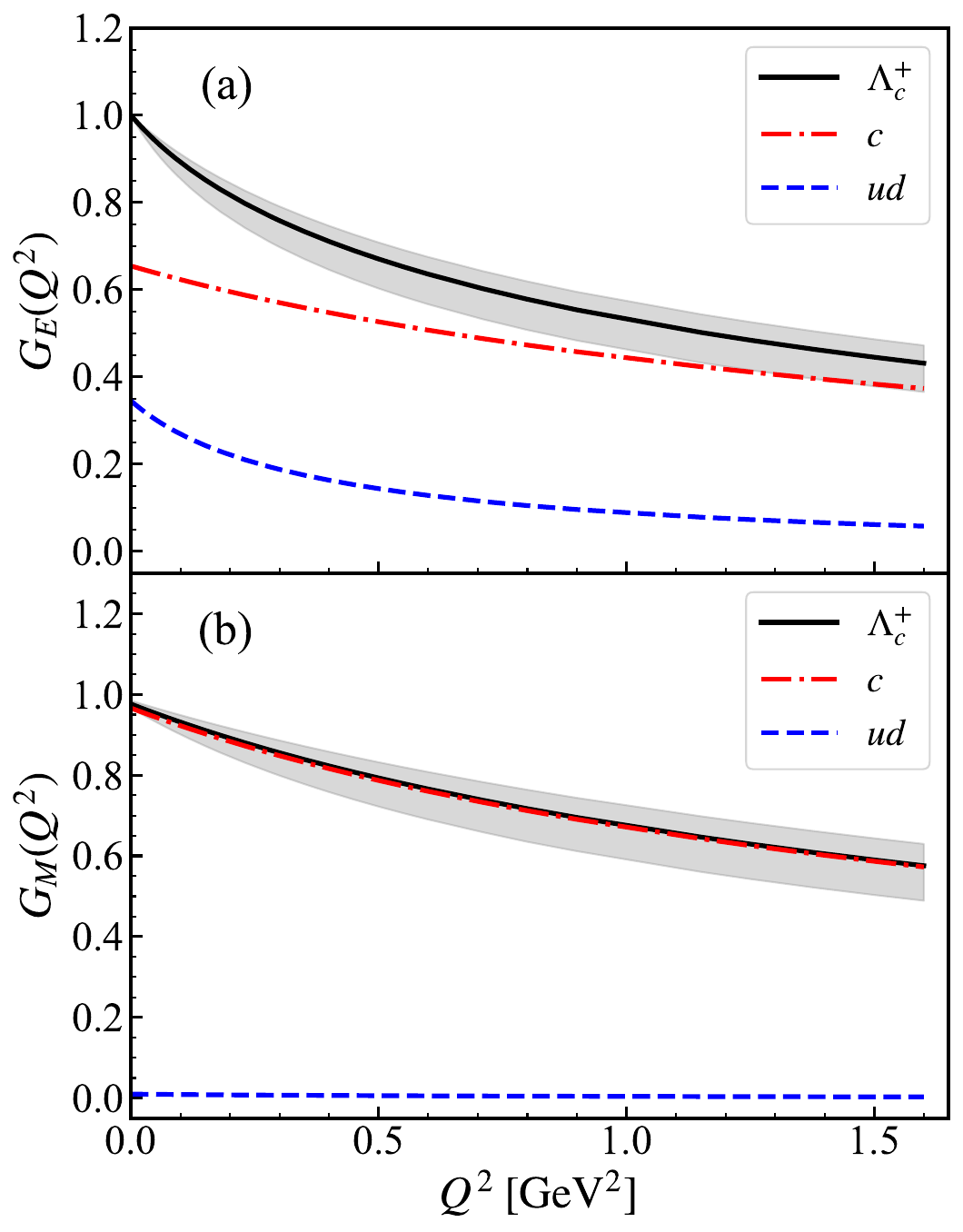}
    \caption{Theoretical calculations for the electric form factor (a) and magnetic form factor (b) of $\Lambda_c^{+}$. The gray band reflects the uncertainty from varying the diquark mass $m_{[qq]}$ between 680 and 700 MeV.} \label{fig:Lambda_c EMFFs}
\end{figure}

While for the $\Lambda_c^+$ baryon, its diquark mass $m_{[qq]}$ is decreased by 170 MeV relative to $\Sigma_c$ based on HQSS, and varied within the range of $160 \sim 180$ MeV to estimate the uncertainty. As shown in Fig.~\ref{fig:Lambda_c EMFFs}, the black solid curve represents the EMFFs under HQSS and the gray band reflects the uncertainty arising from $m_{[qq]}$. It is found that the electric form factors $G_E$ of both $\Lambda_c^+$ and $\Sigma_c^+$ exhibit similar trends, in agreement with the finding of Ref.~\cite{Kim:2018nqf} (see more details in Fig. 3 of that reference). The individual contributions from the charm quark and the $ud$ diquark to $G_E$ also behave similarly to those of $\Sigma_c^+$, reflecting the common electric charge structure of the $cud$ system.

In contrast, the magnetic form factors $G_M$ of $\Lambda_c^+$ and $\Sigma_c^+$ show distinctly different behaviors. As seen in Fig.~\ref{fig:Lambda_c EMFFs}, $G_M$ of $\Lambda_c^+$ is dominated by the charm quark, opposite to the case of $\Sigma_c^+$, where the light diquark dominates. This difference arises because the light diquark in $\Lambda_c^+$ has spin zero and thus contributes little to $G_M$, which can also be directly observed from Eq.~(\ref{eq:magnetic moment}). This also explains why Ref.~\cite{Kim:2018nqf} estimated that $G_M$ of $\Lambda_c^+$ vanishes (leading to zero magnetic moment and magnetic radius), based on their assumption of an infinitely heavy charm quark. Under that assumption, the only source of magnetism would be the light diquark, which is inactive here. 

A direct comparison of $G_M$ for $\Lambda_c^+$ and $\Sigma_c^+$ further reveals that the charm-quark contribution in $\Lambda_c^+$ is significantly larger than that in $\Sigma_c^+$. This can be understood from Eq.~\eqref{eq:magnetic moment}: the magnetic moment contributed by the charm quark in $\Sigma_c^+$ is $-1/3$ of that in $\Lambda_c^+$, a factor originating from the Clebsch–Gordan coefficient when coupling the spin of the charm quark with the axial-vector diquark to form the total spin of $\Sigma_c^+$. Thus, despite the same $cud$ quark content, the different spin configurations of the light quarks lead to completely different contributions from the heavy and light degrees of freedom to $G_M$. Nevertheless, the values of $G_M(0)$ for $\Lambda_c^+$ and $\Sigma_c^+$ are comparable, indicating similar magnetic moments despite their very different internal magnetic structures. However, unlike $\Sigma_c^+$, the $G_M$ of $\Lambda_c^+$ falls off very slowly with increasing $Q^2$, implying a small magnetic radius and a highly localized magnetization distribution, in contrast to the more extended distribution of $\Sigma_c^+$ dominated by the light diquark. These contrasting behaviors highlight the subtle interplay between the heavy quark and light diquark dynamics, and demonstrate that the charm quark with the finite mass plays a crucial role in resolving the magnetic structure of singly charmed baryons.

{To parametrize the $Q^2$ dependence of the form factors and make our results more accessible, we have performed a standard dipole fit. For each baryon and each of the EMFFs, we fit the $Q^2$ dependence using the dipole form,
\begin{align}
    G(Q^2) = \frac{G(0)}{(1 + Q^2/\Lambda^2)^2},
\end{align}
where $G(0)$ and $\Lambda$ are free parameters. For electric form factor of the neutral baryon $\Sigma_c^0$, where $G_E(0)=0$, we use
\begin{align}
    G_E(Q^2) = \frac{a\,Q^2}{(1 + Q^2/\Lambda^2)^2},
\end{align}
with $a$ and $\Lambda$ as the fitted parameters. The fitted parameters and the corresponding adjusted coefficient of determination ($R^2_{\text{adj}}$) are listed in Tab.~\ref{tab:dipole_fit}. The quantity $R^2_{\text{adj}}$ measures the quality of a fit: a value close to 1 indicates an excellent description of the data, while a value near 0 (or negative) indicates a poor fit. We find that all $R^2_{\text{adj}}$ values are close to 1, indicating that the dipole parametrization reproduces the numerical results with high accuracy. We also
compare the dipole functions with the calculated EMFFs in Fig.~\ref{fig:dipole_fit}.

\begin{table}[htbp]
    \centering
    \caption{Fitted parameters of the dipole parametrization. For the neutral $\Sigma_c^0$ electric form factor, the parameters are $a$ and $\Lambda$; for the others, they are $G(0)$ and $\Lambda$. The $R^2_{\text{adj}}$ in the last column is the adjusted coefficient of determination used to measure the quality of each fit.}
    \begin{tabular}{c|ccc}
    \hline\hline
    Electromagnetic form factors & $G(0)$ or $a$ & $\Lambda$ [GeV] & $R^2_{\text{adj}}$ \\
    \hline
    $G_E^{\Sigma_c^{++}}$ & 2.02549 & 1.16542 & 0.99892 \\
    $G_M^{\Sigma_c^{++}}$ & 5.42268 & 0.94889 & 0.99946 \\
    $G_E^{\Sigma_c^{0}}$  & 1.12977 & 1.07312 & 0.95735 \\
    $G_M^{\Sigma_c^{0}}$  & $-3.08593$ & 1.04907 & 0.99908 \\
    $G_E^{\Sigma_c^{+}}$  & 0.98695 & 1.60838 & 0.99913 \\
    $G_M^{\Sigma_c^{+}}$  & 1.16892 & 0.83443 & 0.99991 \\
    $G_E^{\Lambda_c^{+}}$ & 0.98854 & 1.61011 & 0.99934 \\
    $G_M^{\Lambda_c^{+}}$ & 0.97285 & 2.23096 & 0.99994 \\
    \hline\hline
    \end{tabular}
    \label{tab:dipole_fit}
\end{table}

\begin{figure*}[htbp]
    \centering
    \includegraphics[scale=0.42]{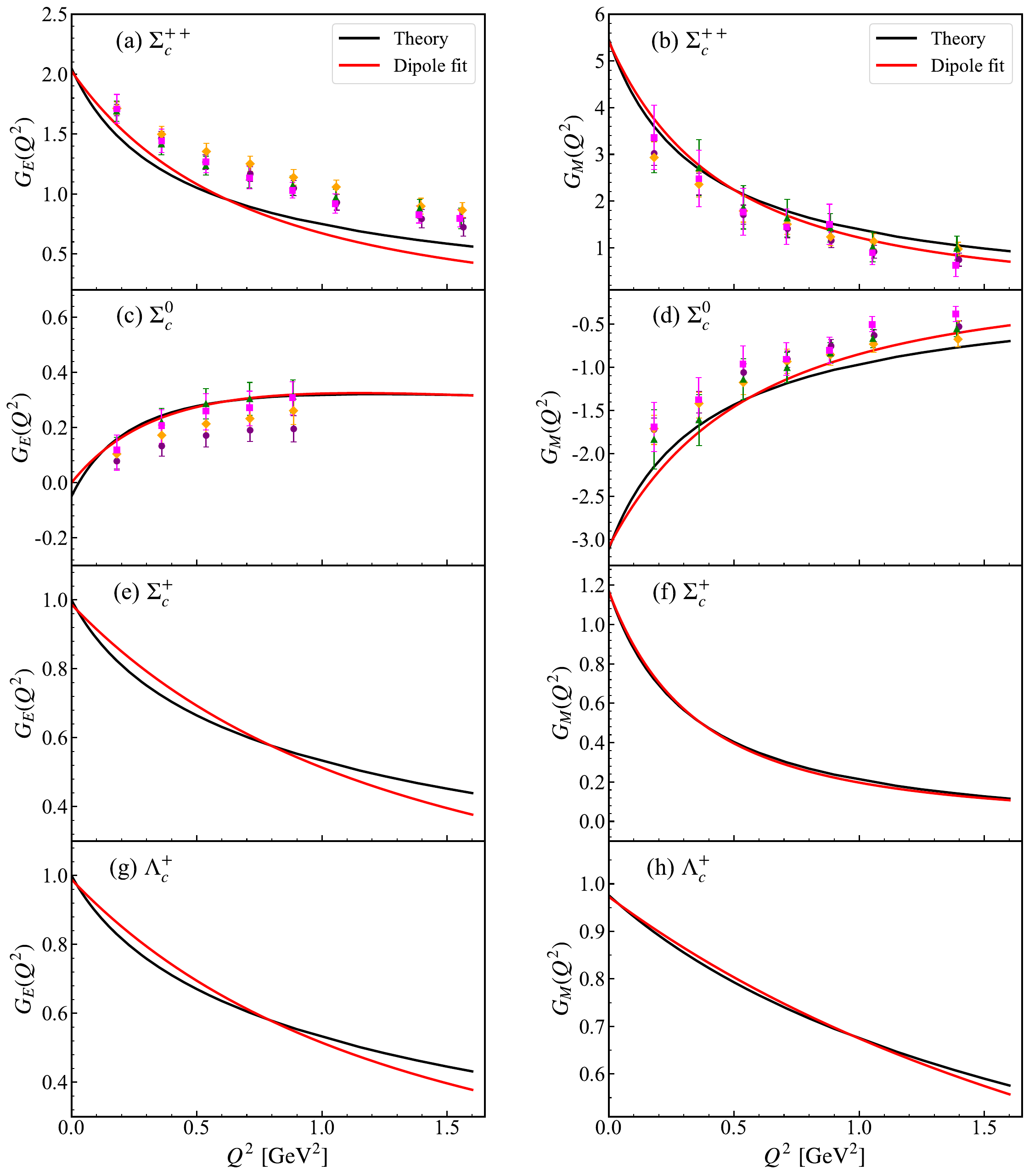}
    \caption{Dipole parametrization of the calculated EMFFs. The black solid curves represent our numerical results, and the red solid curves represent the corresponding dipole fits. The LQCD data, same as in Fig.~\ref{fig:Sigma_c EMFFs}, are shown for comparison where available.}
    \label{fig:dipole_fit}
\end{figure*}

}

\subsection{Electric and magnetic properties of $\Sigma_c$ and $\Lambda_c$}

From the obtained EMFFs, the static EM properties of $\Sigma_c$ and $\Lambda_c^+$ are investigated. Specifically, the magnetic moments as well as the electric and magnetic radii have been extracted. We first present our numerical results for the magnetic moments in Tab.~\ref{tab:magnetic moments}, together with comparisons to various other model calculations~\cite{Can:2013tna,Can:2021ehb,Patel:2025txc,Ozdem:2024brk,Farhadi:2023ucs,Mohan:2022sxm,Kim:2018nqf,Yang:2018uoj,Gandhi:2018lez,Simonis:2018rld,Shi:2018rhk,Wang:2018gpl,Aliev:2015axa,Aliev:2001ig,Bernotas:2013eia,Sharma:2010vv,Faessler:2006ft,Scholl:2003ip,Julia-Diaz:2004yqv,Zhu:1997as,Barik:1983ics,Oh:1991ws}. Although there are no experimental data, theoretical predictions of magnetic moments for $\Sigma_c^{++}$, $\Sigma_c^0$, $\Sigma_c^+$, and $\Lambda_c^+$ mostly cluster around $2\mu_N$, $-\mu_N$, $0.4\mu_N$, and $0.4\mu_N$, respectively. Compared with other theoretical predictions, the hypercentral constituent quark model (hCQM) of Ref.~\cite{Patel:2025txc} is very different. Our results are consistent with these typical values. Notably, our results for $\Sigma_c^{++}$ and $\Sigma_c^0$ fall within the LQCD range obtained via chiral extrapolation using quadratic-form fits. Our results also satisfy the generalized Coleman–Glashow relation arising from the isospin symmetry, $\mu(\Sigma_c^{++})-\mu(\Sigma_c^+)=\mu(\Sigma_c^+)-\mu(\Sigma_c^0)$, as noted in Ref.~\cite{Yang:2018uoj}. These consistencies across different theoretical constraints strengthen the confidence in our predictions for the yet-unmeasured EM properties of these baryons.

\begin{table*}[htpb]
    \centering
    \caption{The magnetic moments of the singly charmed baryons $\Sigma_c$ and $\Lambda_c^+$ are presented in unit of $\mu_N$. Our results are listed in the second column of the upper half of the table, and are compared with other theoretical predictions~\cite{Can:2013tna,Can:2021ehb,Patel:2025txc,Ozdem:2024brk,Farhadi:2023ucs,Mohan:2022sxm,Kim:2018nqf,Yang:2018uoj,Gandhi:2018lez,Simonis:2018rld,Shi:2018rhk,Wang:2018gpl,Aliev:2015axa,Aliev:2001ig,Bernotas:2013eia,Sharma:2010vv,Faessler:2006ft,Scholl:2003ip,Julia-Diaz:2004yqv,Zhu:1997as,Barik:1983ics,Oh:1991ws}. Values in parentheses denote the errors, while values in the square brackets correspond to another scenarios adopted in the respective references.}
    \label{tab:magnetic moments}
    \scalebox{0.9}{
    \begin{tabular}{ccccccccccccccc}
    \hline\hline &\multirow{2}{*}{Ours} &LQCD &hCQM
    &LCQSR&QDM&EMS & \multicolumn{2}{c}{SQCS\cite{Mohan:2022sxm}}&$\chi$QSM&$\chi$QSM&hCQM &BM\\\cline{8-9}
    Baryons & &\cite{Can:2013tna, Can:2021ehb}&\cite{Patel:2025txc}&\cite{Ozdem:2024brk}&\cite{Farhadi:2023ucs}&\cite{Mohan:2022sxm}    &$z=0.021$ &$z=0.155$ &\cite{Kim:2018nqf}&\cite{Yang:2018uoj}&\cite{Gandhi:2018lez}&\cite{Simonis:2018rld}\\
    \hline
    $\Sigma_c^{++}(uuc)$  &2.088&2.220(505)&2.166&2.02(18)&1.782&2.095&2.365&2.622&1.58[1.60]&2.15(10)&1.831&2.28\\
    $\Sigma_c^+(udc)$ &0.448 &-&1.216&0.50(5)&0.364&0.432&0.534&0.818&0.39[0.33]&0.46(3)&0.380&0.487\\
    $\Sigma_c^0(ddc)$ &$-1.191$
    &$-1.073(269)$&0.135&$-1.01(9)$&$-1.061$&$-1.234$ &$-1.299$ &$-0.987$&$-0.79[-0.94]$&$-1.24(5)$&$-1.091$&$-1.31$\\
    $\Lambda_c^+(udc)$  &0.400  -&-&0.408&0.46(9)&0.374&0.380&0.384&0.410&0&0&0.421&0.335\\
    \hline
    &B$\chi$PT&HB$\chi$PT&LCQSR&BM&$\chi$CQM&RQM&SM&RQM&QSR&PM&SM\\
    Baryons&\cite{Shi:2018rhk}&\cite{Wang:2018gpl}&\cite{Aliev:2015axa,Aliev:2001ig}&\cite{Bernotas:2013eia}&\cite{Sharma:2010vv}&\cite{Faessler:2006ft}&\cite{Scholl:2003ip}&\cite{Julia-Diaz:2004yqv}&\cite{Zhu:1997as}&\cite{Barik:1983ics}&\cite{Oh:1991ws}\\
    \hline
    $\Sigma_c^{++}(uuc)$ &2.00 &1.50&2.4(5)&1.679&2.20&1.76&2.45&$0.9-3.07$&2.1(3)&2.359&1.95  \\
    $\Sigma_c^+(udc)$ &0.46 &0.30&0.50(15)&0.318&0.30&0.36&0.25&$0.11-0.65$ &0.23(3)&0.505&0.41\\
    $\Sigma_c^0(ddc)$ &$-1.08$ &$-0.91$&$-1.50(35)$&$-1.043$&$-1.60$&$-1.04$&$-1.96$&$-(0.67-1.78)$&-1.6(2)&$-1.349$&$-1.1$\\
    $\Lambda_c^+(udc)$  &0.24  &0.24&0.40(5)&0.411&0.392&0.42&0.12&$0.39-0.52$&0.15(5)&0.34&-\\
    \hline\hline
    \end{tabular}}
\end{table*}

In addition, the electric and magnetic radii extracted from Eq.~(\ref{eq:radii}) are also important quantities, as they are determined by the behavior of the EMFFs near $Q^2=0$ and provide direct information on the sizes of the baryons and listed in Tab.~\ref{tab:radii}. Our results show that for both $\Sigma_c$ and $\Lambda_c^+$, the electric radii are systematically smaller than the corresponding magnetic radii, suggesting that the magnetization distributions are more spatially extended. For comparison, we also list the predictions from a few other theoretical approaches. Our results for the electric radii are larger than the LQCD estimates but are more consistent with those from the $\chi$QSM. In all cases, the electric radii of the singly charmed baryons are noticeably smaller than that of the proton, whose latest experimental value is $\langle r^2\rangle_E^p=0.707\ \text{fm}^2$~\cite{ParticleDataGroup:2024cfk}, reflecting their more compact charge distributions. Moreover, the relation $\langle r^2\rangle_E^{\Sigma_c^{++}}>\langle r^2\rangle_E^{\Sigma_c^+}$ can be naturally understood from their charge difference: the doubly $\Sigma_c^{++}$ carries twice the charge of $\Sigma_c^+$, leading to a larger electric radius. Notably, the electric radius of $\Lambda_c^+$ is even smaller than that of $\Sigma_c^+$, indicating a more compact charge distribution.

The magnetic radii of $\Sigma_c$ and $\Lambda_c^+$ are also listed in Tab.~\ref{tab:radii}. Among the three $\Sigma_c$ baryons, $\Sigma_c^{+}$ has the largest magnetic radius. The magnetic radii of the $\Sigma_c$ are found to be comparable to that of the proton, with the latest experimental value $\langle r^2\rangle_M^p=0.724\ \text{fm}^2$~\cite{ParticleDataGroup:2024cfk}, and are not significantly different from the LQCD and $\chi$QSM results. Furthermore, we find that both the electric and magnetic radii of $\Lambda_c^+$ are considerably smaller than those of $\Sigma_c^+$, indicating that $\Lambda_c^+$ is more compact than $\Sigma_c^+$. This is consistent with the fact that the scalar diquark ($[qq]$) in $\Lambda_c^+$ forms a more tightly bound configuration than the axial-vector diquark in $\Sigma_c$ ($\{qq\}$), as reflected simply by the relation $2m_q-m_{[qq]}>2m_q-m_{\{qq\}}$. Together, these results provide a coherent picture of the electric charge and magnetization distributions in singly charmed baryons and highlight the sensitivity of the EM radii to both the spin configuration and the charge of the system.

\begin{table}[htbp]
    \centering
    \caption{The electric and magnetic radii (in fm$^2$) of $\Sigma_c$ and $\Lambda_c^+$ obtained from different approaches. In the $\chi$QSM results~\cite{Kim:2018nqf}, the values outside and inside the square brackets in the second-to-last column correspond to strange current quark masses of $m_s=0$ MeV and $174$ MeV, respectively. The last column presents the RQM results~\cite{Julia-Diaz:2004yqv}, where the three values in square brackets are obtained using three different forms of kinematics (see more details in that reference): instant, point, and front forms.}
    \label{tab:radii}
    \scalebox{0.9}{
    \begin{tabular}{ccccccccccccc}
         \hline\hline
        &\multirow{2}{*}{} &{Ours}&LQCD~\cite{Can:2013tna,Can:2021ehb}&$\chi$QSM~\cite{Kim:2018nqf}&RQM~\cite{Julia-Diaz:2004yqv}\\
        \hline
        \multirow{2}{*}{$\Sigma_c^{++}$} &${\langle r_{E}^2\rangle}$&0.541&0.234(37)&0.60[0.59]&[1.7,0.4,1.4]  \\
        &${\langle r_{M}^2\rangle}$&0.674&0.696(153)&0.62[0.62]&-  \\
        \hline
        \multirow{2}{*}{$\Sigma_c^0$}&$\langle r_{E}^2\rangle$&$-0.438$&-&$-0.60[-0.66]$&$[-0.7,-0.0,-0.6]$\\
        &${\langle r_{M}^2\rangle}$&0.605&0.650(126)&0.62[0.78]&-\\
        \hline
        \multirow{2}{*}{$\Sigma_c^+$}&${\langle r_{E}^2\rangle}$&0.322&-&0.30[0.27]&[0.5,0.2,0.4]  \\
        &${\langle r_{M}^2\rangle}$&0.766&-&0.62[0.40]&- \\
        \hline
        \multirow{2}{*}{$\Lambda_c^{+}$}&${\langle r_{E}^2\rangle}$ &0.312&-&0.26[0.24]&[0.5,0.2,0.4]  \\
        &${\langle r_{M}^2\rangle}$&0.114&-&-&-   \\
        \hline\hline
    \end{tabular}}
\end{table}

{
\subsection{Sensitivity analysis of model parameters}

In this subsection, we examine the sensitivity of the numerical results to the model parameters. As discussed in Sec.~\ref{sec:parameters}, these parameters have been fixed by fitting to LQCD data or adjusted within the reasonable range determined in other literature. It is therefore important to assess whether our conclusions are sensitive to moderate variations of these parameters. 

We first investigate the dependence on the cutoff parameter $m_R$, which regulates the loop integrals via the vertex function $\Xi(p_1,p_2)$ in Eq.~(\ref{eq:scalar function}). We vary $m_R$ within the range $2.8$--$3.4$~GeV around its central value $m_R = 3.1$~GeV, corresponding to about $\pm 10\%$. As an illustrative example, we show the form factors $G_E$ and $G_M$ of $\Sigma_c^0$ for the three different values of $m_R$ in Fig.~\ref{fig:sensitivity} (a) and (b). The three curves are almost overlapping, except for the $G_E$ of $\Sigma_c^0$. To quantify this further, we list in Tab.~\ref{tab:mR_sensitivity} the magnetic moments and electric and magnetic radii extracted from these three sets of form factors. The relative variations are all slight, confirming that our results are not sensitive to the specific choice of $m_R$, and that the cutoff scheme is reliable in this work.


\begin{figure*}[htbp]
    \centering
    \includegraphics[scale=0.42]{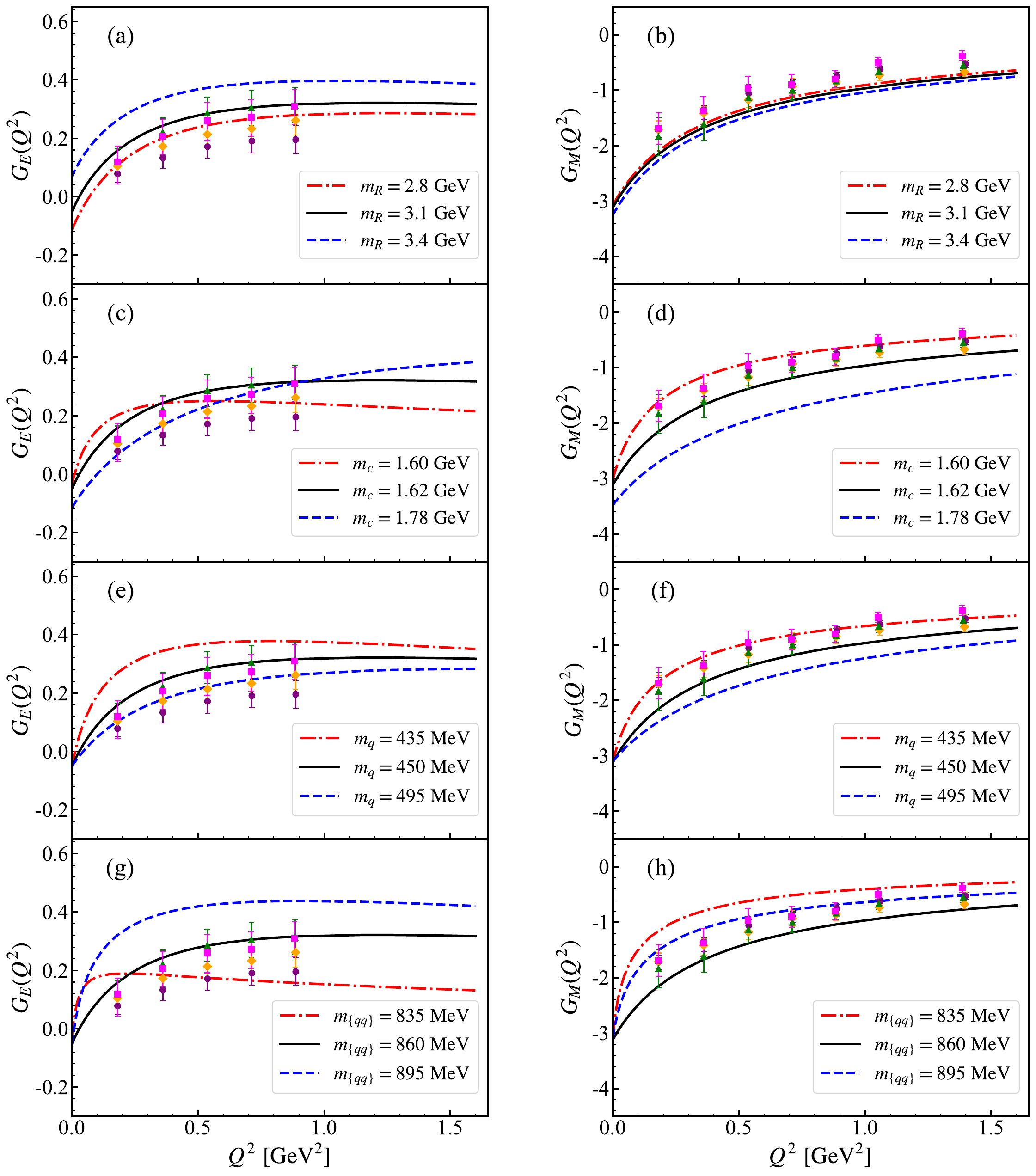}
    \caption{Sensitivity of the EMFFs of $\Sigma_c^0$ to the model parameters. The left and right panels show $G_E$ and $G_M$, respectively. From top to bottom, the curves correspond to different values of $m_R$, $m_c$, $m_q$, and $m_{\{qq\}}$, as indicated in the legends.}
    \label{fig:sensitivity}
\end{figure*}

\begin{table}[htbp]
    \centering
    \caption{Magnetic moments (in $\mu_N$) and electric and magnetic radii (in fm$^2$) of $\Sigma_c$ and $\Lambda_c^+$ for different values of $m_R$.}
    \begin{tabular}{c|c|ccc}
        \hline\hline
        \multirow{2}{*}{Baryons} & \multirow{2}{*}{Observables} & \multicolumn{3}{c}{$m_R$~[GeV]} \\
        \cline{3-5}
        & & 2.8 & 3.1 & 3.4 \\
        \hline
        $\Sigma_c^{++}$ & $\mu$ & 2.055 & 2.088 & 2.142 \\
        & $\langle r_E^2\rangle$ & 0.566 & 0.541 & 0.544 \\
        & $\langle r_M^2\rangle$ & 0.699 & 0.674 & 0.650 \\
        \hline
        $\Sigma_c^0$ & $\mu$ & $-1.174$ & $-1.191$ & $-1.221$ \\
        & $\langle r_E^2\rangle$ & $-0.464$ & $-0.438$ & $-0.440$ \\
        & $\langle r_M^2\rangle$ & 0.625 & 0.605 & 0.584 \\
        \hline
        $\Sigma_c^+$ & $\mu$ & 0.440 & 0.448 & 0.461 \\
        & $\langle r_E^2\rangle$ & 0.334 & 0.322 & 0.324 \\
        & $\langle r_M^2\rangle$ & 0.799 & 0.766 & 0.738 \\
        \hline
        $\Lambda_c^+$ & $\mu$ & 0.398 & 0.400 & 0.403 \\
        & $\langle r_E^2\rangle$ & 0.325 & 0.312 & 0.302 \\
        & $\langle r_M^2\rangle$ & 0.119 & 0.114 & 0.110 \\
        \hline\hline
    \end{tabular}
    \label{tab:mR_sensitivity}
\end{table}

We next examine the sensitivity to the constituent quark and diquark masses ($m_c,~m_q,~m_{D}$ with $q=u,d$ and $D=\{qq\},[qq]$). As discussed in Sec.~\ref{sec:parameters}, these masses differ in different phenomenological implementations. To quantify the associated theoretical uncertainties, we vary one mass parameter at a time while keeping the other two fixed at their central values in Tab.~\ref{tab:mass para}. Each parameter is varied by $\pm 10\%$ around its central value. However, not all of these variations are physically allowed, since the input masses must satisfy the bound-state conditions $M < m_c + m_D$ and $m_D < 2m_q$. When a $\pm 10\%$ variation would violate these conditions, we truncate the variation at the bound-state boundary and use the corresponding limiting value. The resulting effective ranges are indicated in the legends of Fig.~\ref{fig:sensitivity}.

To save space, we also use $\Sigma_c^0$ as an illustrative example. As shown in Fig.~\ref{fig:sensitivity} (c)--(h), almost all form factors exhibit noticeable sensitivity to the three mass parameters. However, it is evident that the values of the form factors at $Q^2=0$ are insensitive to variations of the mass parameters, whereas the slopes at $Q^2=0$ show much larger discrepancies. This demonstrates that the magnetic moments are relatively stable observables when varying the mass parameters. For completeness, the magnetic moments obtained with each parameter set are collected in Tab.~\ref{tab:mass_sensitivity}. 

The EM radii, by contrast, show a more pronounced sensitivity to the mass parameters than the magnetic moments. Since these variations are large and do not provide a meaningful reference for comparison, we do not list them individually radius in the table and instead focus on the magnetic moments, which are the more robust observables. Furthermore, although not shown in Fig.~\ref{fig:sensitivity}, the form factors of $\Lambda_c^+$ are remarkably stable against variation of the light quark $m_q$. This is consistent with our earlier finding that the magnetic form factor of $\Lambda_c^+$ is governed almost entirely by the charm quark. 

Although the line shapes of the EMFFs are sensitive to the mass parameters, we would like to emphasize that, within the quark-diquark model, it is possible to describe the EMFFs of singly charmed baryons $\Sigma_c$ and $\Lambda_c^+$ using physically acceptable parameters. This is because the constituent quark and diquark masses are effective parameters that vary across different phenomenological implementations. Furthermore, the central values in Tab.~\ref{tab:mass para} are chosen to give a best quantitative description with LQCD results, ensuring that both the magnetic moments and the EM radii are comparable to other theoretical results.


\begin{table*}[htbp]
    \centering
    \caption{Magnetic moments (in $\mu_N$) of $\Sigma_c$ and $\Lambda_c^+$ with each mass parameter set. In each test, one mass parameter is varied while the other two are held fixed at their central values. The middle column of each group corresponds to the central-value result, which serves as the reference for comparison.}
    \begin{tabular}{c|ccc|ccc|ccc}
        \hline\hline
        \multirow{2}{*}{Baryons} &\multicolumn{3}{c|}{$m_c$~[GeV]} &\multicolumn{3}{c|}{$m_q$~[MeV]} &\multicolumn{3}{c}{$m_{\{qq\}}$~[MeV]}\\
        \cline{2-10}
        & 1.60 &1.62 &1.78 &435 &450 &495 &835 &860 &895\\
        \hline
        $\Sigma_c^{++}$ &1.967 &2.088 &2.493 &2.078 &2.088 &2.076 &1.973 &2.088 &2.095\\
        $\Sigma_c^0$ &$-1.154$ &$-1.191$ &$-1.330$ &$-1.187$ &$-1.192$ &$-1.186$ &$-1.165$ &$-1.191$ &$-1.177$\\
        $\Sigma_c^+$ &0.407 &0.448 &0.581 &0.446 &0.448 &0.445 &0.404 &0.448 &0.459\\
        \hline
        \multirow{2}{*}{Baryons} &\multicolumn{3}{c|}{$m_c$~[GeV]} &\multicolumn{3}{c|}{$m_q$~[MeV]} &\multicolumn{3}{c}{$m_{[qq]}$~[MeV]}\\
        \cline{2-10}
        & 1.6 &1.62 &1.78 &405 &450 &495 &675 &690 &755\\
        \hline
        $\Lambda^+$ &0.396 &0.400 &0.400 &0.400 &0.400 &0.400 &0.394 &0.400 &0.418\\
        \hline\hline
    \end{tabular}
    \label{tab:mass_sensitivity}
\end{table*}

}

\subsection{Total cross sections for the $e^+e^-\to \Sigma_c\bar{\Sigma_c}$ reactions}

Although the form factors are real functions of $Q^2$ in the spacelike region and complex functions of $q^2$ in the timelike region, they become real in the large-$q^2$ limit, as required by the Schwarz reflection principle~\cite{Pacetti:2014jai}. 
To connect the EMFFs between the two regions, one may employ the following empirical asymptotic relation,
\begin{equation}
 G_{E,M}^{\text{TL}}(q^2)\simeq G_{E,M}^{\text{SL}}(-q^2 + xM^2), \label{eq:relation}
 \end{equation}
which follows from the unitarity of the EM current and the Phragm\'en-Lindel\"of theorem~\cite{Pacetti:2014jai}.
In Eq.~\eqref{eq:relation} and the following equations, $\text{TL}$ and $\text{SL}$ in the superscript respectively represents the timelike and spacelike regions.
The parameter $x$, introduced in Ref.~\cite{Ramalho:2019koj}, serves to test the validity of this asymptotic behavior and to locate the onset of the finite-$q^2$ region where it becomes a reasonably good approximation. In practice, $x=0$ and $x=4$ are taken as the lower and upper limits, and the optimal value can be determined by comparing with the available data. The $x$-dependence naturally vanishes at very large $q^2$, consistent with the asymptotic identity between these two regions. The Ref.~\cite{Ramalho:2019koj} extended the spacelike EMFFs of hyperons to the timelike region and compared the estimated effective form factors with experimental data (see Figs. 1--3 therein), confirming the validity of the asymptotic relation at $x=2$ in the hyperon systems. 

In the timelike region, for the process $e^+e^-\to B\bar{B}$ ($B$ stands for a baryon and $\bar{B}$ is an antibaryon), its total cross section is given by~\cite{Korner:1976hv,BaBar:2005pon}
\begin{align}
    \begin{split}
        \sigma(q^2)=\frac{4\pi\alpha^2\beta C}{3q^2}\Big(1+\frac{1}{2\tau}\Big)|G_{\text{eff}}(q^2)|^2,
    \end{split}
\end{align}
where $\alpha$ is the fine structure constant, $\tau=q^2/(4M^2)$, $\beta=\sqrt{1-1/\tau}$, and $C=y/(1-e^{-y})$ is the Coulomb correction factor with $y=\pi\alpha M/(2\sqrt{\tau-1})$. The effective form factor $|G_{\text{eff}}(q^2)|$ for the baryon $B$ is expressed in terms of the timelike EMFFs $G_E^{TL}(q^2)$ and $G_M^{TL}(q^2)$ as,
\begin{align}
    \begin{split}
        |G_{\text{eff}}(q^2)|&=\sqrt{\frac{2\tau|G_M^{TL}(q^2)|^2+|G_E^{TL}(q^2)|^2}{1+2\tau}}.
    \end{split}
\end{align}

\begin{figure*}[htbp]
    \centering
    \includegraphics[scale=0.42]{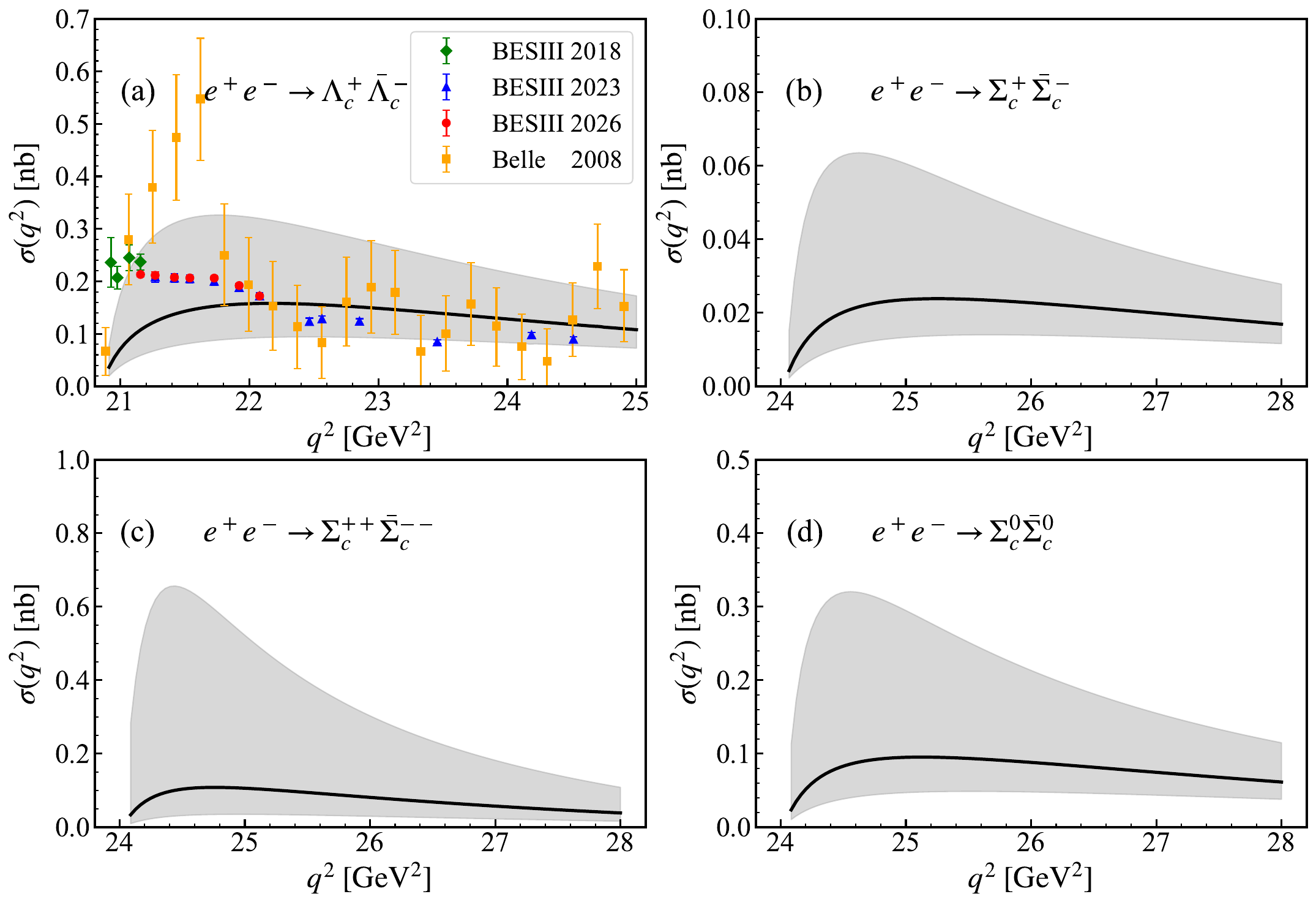}
\caption{Theoretical calculations of the total cross sections for $e^+ e^- \to \Lambda^+_c \bar{\Lambda}^-_c$ and $\Sigma_c \bar{\Sigma}_c$ reactions. The solid curve in panel (a) shows the total cross section for $e^+e^-\to \Lambda_c^+\bar{\Lambda}_c^-$ obtained from the asymptotic relation, which has been fitted to the BESIII data~\cite{BESIII:2023rwv,BESIII:2026qbp}, resulting in $x=3.5$. The gray band corresponds to a $10\%$ variation of the parameter $x$. Panels (b)-(d) present the predicted total cross sections for $e^+e^-\to\Sigma_c\bar{\Sigma}_c$ using the same asymptotic relation.} \label{fig:cross section}
\end{figure*}

With Eq.~(\ref{eq:relation}) we can easily obtain the timelike EMFFs of $\Lambda_c^+$ and thereby calculate the total cross section for $e^+e^-\to\Lambda_c^+\bar{\Lambda}_c^-$. By fitting to the BESIII experimental data, we determine $x= 3.5$ as shown in Fig.~\ref{fig:cross section} (a) (black solid curve), which differs from that of Ref.~\cite{Ramalho:2019koj}. However, this difference is understandable, given the significantly different $q^2$ regions of the respective reactions. We also vary $x$ by its $10\%$ ($x = 3.5 \pm 0.35$) to estimate the uncertainty associated with this empirical relation, corresponding to the gray band in the figure. Given the similar masses of $\Lambda_c^+$ and $\Sigma_c$, the asymptotic relation determined from the $\Lambda_c^+$ data can guide future $e^+e^-\to\Sigma_c\bar{\Sigma}_c$  experiments. As shown in Fig.~\ref{fig:cross section}, the estimated total cross section of $e^+ e^- \to \Sigma^+_c \bar{\Sigma}^-_c$ is one order smaller than others. It is expected that these model calculations can be tested by future experimental measurements.

It should be noted that charmonium states also contribute to these reactions, as reported in Refs.~\cite{Chen:2023oqs,Chen:2024luh}. Here, by employing the relation in Eq.~\eqref{eq:relation}, we aim to provide an estimate of the total cross sections for these processes.

\section{Summary}     \label{sec-summary}

In this work, we have systematically investigated the spacelike EMFFs of the singly charmed baryons, $\Sigma_c$ ($\Sigma_c^{++},\Sigma_c^+,\Sigma_c^0$) and $\Lambda_c^+$, using a covariant quark-diquark model and focusing on a comparison of the results with the LQCD data. Guided by HQSS, the two light quarks in the singly charmed baryon are treated as an effective diquark, which naturally reduces the degrees of freedom while preserving the essential QCD dynamics. The quark–diquark picture also provides a natural framework for revealing how the heavy quark and light diquark separately influence the EM properties of the singly charmed baryons.

Our analysis shows that the electric form factors of both $\Sigma_c$ and $\Lambda_c^+$ fall off significantly slower with $Q^2$ than that of the proton, leading to much smaller electric radii. This confirms that the presence of a heavy charm quark acts as a localized core, effectively shrinking the charge distribution. The magnetic properties, however, exhibit a more subtle behavior dictated by the spin configuration of the light diquark. For $\Sigma_c$ baryons, the magnetic form factors are dominated by the light degree of freedom, and the resulting magnetic radii are found to be comparable to that of the proton, indicating an extended magnetization distribution from the light sector. Furthermore, the opposite signs observed between the heavy-quark and light-diquark contributions to the magnetic moments imply a predominantly antialigned spin configuration. 

In contrast, for the lightest charmed baryon $\Lambda_c^+$, the scalar diquark has spin zero and thus contributes negligibly to the magnetism. As a result, the magnetic form factor is governed by the charm quark, leading to a magnetic radius much smaller than that of the proton and reflecting a highly localized magnetization distribution. Notably, the magnetic properties of $\Lambda_c^+$ present an exceptional case where the infinite heavy-quark mass limit fails: in this limit the magnetic form factor would vanish, whereas our finite-mass framework predicts a nonzero, slowly decreasing $G_M(Q^2)$ driven almost by the charm quark. This highlights the necessity to go beyond the static approximation when describing the magnetic structure of spin-0 diquark systems. 

{We further examine the sensitivity of our results to these model parameters. Varying the cutoff $m_R$ over 2.8--3.4~GeV leads to slight changes in the theoretical results, except for $G_E$ of the neutral $\Sigma_c^0$, and shifts the magnetic moments and radii by only a few percent. Varying the constituent masses $m_c$, $m_q$, and $m_D$ within physically allowed ranges produces visible changes in the form factors. However, the values at $Q^2=0$ which correspond to the magnetic moments, remain relatively stable, while the electromagnetic radii show a stronger dependence. In particular, the results for $\Lambda_c^+$ are almost insensitive to $m_q$, as expected since its magnetism is governed by the charm quark.
}

Finally, based on the experimental measurements of the total cross section for the $e^+e^-\to \Lambda_c^+\bar{\Lambda}_c^-$ reaction, we determine the asymptotic relation connecting the timelike and spacelike EMFFs and use it to extend the $\Sigma_c$ EMFFs to the timelike region, thereby estimating the total cross section for these $e^+e^-\to\Sigma_c\bar{\Sigma}_c$ reactions.

In the conclusion, the quark-diquark model provides a coherent and economical description of the EM properties of singly charmed baryons. The good agreement with LQCD data and other phenomenological models supports the reliability of our predictions for the yet-unmeasured spacelike EMFFs, which may be tested in future high-precision experiments at charm factories.

\begin{acknowledgments}

This work is partly supported by the National Key R\&D Program of China under Grant No. 2023YFA1606703, and by the National Natural Science Foundation of China under Grants No. 12575094, No. 12435007, No. 12361141819, and No. 12447121. It is also supported by the Gansu Province Postdoctoral Foundation.

\end{acknowledgments}

\bibliography{ref.bib}

@article{Shifman:2024kfj,
    author = "Shifman, Mikhail",
    title = "{QCD chemistry: Remarks on diquarks}",
    eprint = "2412.05440",
    archivePrefix = "arXiv",
    primaryClass = "hep-ph",
    reportNumber = "FTPI-MINN-24-25, UMN-TH-4405/24",
    doi = "10.1016/j.nuclphysbps.2024.10.007",
    journal = "Nucl. Part. Phys. Proc.",
    volume = "347",
    pages = "86--89",
    year = "2024"
}

@article{Cotogno:2019vjb,
    author = "Cotogno, Sabrina and Lorc{\'e}, C{\'e}dric and Lowdon, Peter and Morales, Manuel",
    title = "{Covariant multipole expansion of local currents for massive states of any spin}",
    eprint = "1912.08749",
    archivePrefix = "arXiv",
    primaryClass = "hep-ph",
    doi = "10.1103/PhysRevD.101.056016",
    journal = "Phys. Rev. D",
    volume = "101",
    number = "5",
    pages = "056016",
    year = "2020"
}

@article{Ernst:1960zza,
    author = "Ernst, F. J. and Sachs, R. G. and Wali, K. C.",
    title = "{Electromagnetic form factors of the nucleon}",
    doi = "10.1103/PhysRev.119.1105",
    journal = "Phys. Rev.",
    volume = "119",
    pages = "1105--1114",
    year = "1960"
}

@article{Frederico:2009fk,
    author = "Frederico, T. and Pace, E. and Pasquini, B. and Salme, G.",
    title = "{Pion Generalized Parton Distributions with covariant and Light-front constituent quark models}",
    eprint = "0907.5566",
    archivePrefix = "arXiv",
    primaryClass = "hep-ph",
    doi = "10.1103/PhysRevD.80.054021",
    journal = "Phys. Rev. D",
    volume = "80",
    pages = "054021",
    year = "2009"
}

@article{Scadron:1968zz,
    author = "Scadron, Michael D.",
    title = "{Covariant Propagators and Vertex Functions for Any Spin}",
    doi = "10.1103/PhysRev.165.1640",
    journal = "Phys. Rev.",
    volume = "165",
    pages = "1640--1647",
    year = "1968"
}

@article{Guo:2024pti,
    author = "Guo, Di and Yang, Qin-He and Dai, Ling-Yun",
    title = "{Study of the timelike electromagnetic form factors of the $\Lambda_c$}",
    eprint = "2404.06191",
    archivePrefix = "arXiv",
    primaryClass = "hep-ph",
    doi = "10.1103/PhysRevD.109.114005",
    journal = "Phys. Rev. D",
    volume = "109",
    number = "11",
    pages = "114005",
    year = "2024"
}

@article{Milstein:2022bfg,
    author = "Milstein, A. I. and Salnikov, S. G.",
    title = "{Final-state interaction in the process $e^+e^-\to\Lambda_c\bar{\Lambda}_c$}",
    eprint = "2201.07450",
    archivePrefix = "arXiv",
    primaryClass = "hep-ph",
    doi = "10.1103/PhysRevD.105.074002",
    journal = "Phys. Rev. D",
    volume = "105",
    number = "7",
    pages = "074002",
    year = "2022"
}

@article{Salnikov:2023qnn,
    author = "Salnikov, S. G. and Milstein, A. I.",
    title = "{Near-threshold resonance in $e^+e^-\to\Lambda_c\bar{\Lambda}_c$ process}",
    eprint = "2309.17018",
    archivePrefix = "arXiv",
    primaryClass = "hep-ph",
    doi = "10.1103/PhysRevD.108.L071505",
    journal = "Phys. Rev. D",
    volume = "108",
    number = "7",
    pages = "L071505",
    year = "2023"
}

@article{BESIII:2023rwv,
    author = "Ablikim, Medina and others",
    collaboration = "BESIII",
    title = "{Measurement of Energy-Dependent Pair-Production Cross Section and Electromagnetic Form Factors of a Charmed Baryon}",
    eprint = "2307.07316",
    archivePrefix = "arXiv",
    primaryClass = "hep-ex",
    doi = "10.1103/PhysRevLett.131.191901",
    journal = "Phys. Rev. Lett.",
    volume = "131",
    number = "19",
    pages = "191901",
    year = "2023"
}

@article{Dai:2024lau,
    author = "Dai, Ling-Yun and Haidenbauer, Johann and Mei{\ss}ner, Ulf-G.",
    title = "{Electromagnetic Form Factors of Hyperons in the Timelike Region: A Short Review}",
    eprint = "2412.07543",
    archivePrefix = "arXiv",
    primaryClass = "hep-ph",
    doi = "10.1088/0256-307X/42/3/030202",
    journal = "Chin. Phys. Lett.",
    volume = "42",
    number = "3",
    pages = "030202",
    year = "2025"
}

@article{Chen:2023oqs,
    author = "Chen, Cheng and Yan, Bing and Xie, Ju-Jun",
    title = "{Cross Sections and the Electromagnetic Form Factors within the Extended Vector Meson Dominance Model}",
    eprint = "2312.16753",
    archivePrefix = "arXiv",
    primaryClass = "hep-ph",
    doi = "10.1088/0256-307X/41/2/021302",
    journal = "Chin. Phys. Lett.",
    volume = "41",
    number = "2",
    pages = "021302",
    year = "2024"
}

@article{Chen:2024luh,
    author = "Chen, Cheng and Yan, Bing and Xie, Ju-Jun",
    title = "{The electromagnetic form factors and spin polarization of $\Lambda_c^+$ in the process $e^+e^-\to\Lambda_c^+\bar{\Lambda}_c^-$}",
    eprint = "2407.19445",
    archivePrefix = "arXiv",
    primaryClass = "hep-ph",
    doi = "10.1088/1674-1137/ad9259",
    journal = "Chin. Phys. C",
    volume = "49",
    number = "2",
    pages = "023102",
    year = "2025"
}

@article{BESIII:2017kqg,
    author = "Ablikim, Medina and others",
    collaboration = "BESIII",
    title = "{Precision measurement of the $e^{+}e^{-}~\rightarrow~\Lambda_{c}^{+} \bar{\Lambda}_{c}^{-}$ cross section near threshold}",
    eprint = "1710.00150",
    archivePrefix = "arXiv",
    primaryClass = "hep-ex",
    doi = "10.1103/PhysRevLett.120.132001",
    journal = "Phys. Rev. Lett.",
    volume = "120",
    number = "13",
    pages = "132001",
    year = "2018"
}

@article{Wan:2021ncg,
    author = "Wan, Junyao and Yang, Yongliang and Lu, Zhun",
    title = "{The electromagnetic form factors of $\Lambda _c$ hyperon in the vector meson dominance model}",
    eprint = "2102.03092",
    archivePrefix = "arXiv",
    primaryClass = "hep-ph",
    doi = "10.1140/epjp/s13360-021-01919-6",
    journal = "Eur. Phys. J. Plus",
    volume = "136",
    number = "9",
    pages = "949",
    year = "2021"
}

@article{BESIII:2026qbp,
    author = "Ablikim, Medina and others",
    collaboration = "BESIII",
    title = "{Measurements of the absolute branching fractions of the $\Lambda_{c}^{+}$ hadronic decays}",
    eprint = "2601.01503",
    archivePrefix = "arXiv",
    primaryClass = "hep-ex",
    journal = "",
    month = "1",
    year = "2026"
}

@article{Aliev:2001ig,
    author = "Aliev, T. M. and Ozpineci, A. and Savci, M.",
    title = "{The Magnetic moments of $\Lambda_b$ and $\Lambda_c$ baryons in light cone QCD sum rules}",
    eprint = "hep-ph/0107196",
    archivePrefix = "arXiv",
    reportNumber = "METU-PHYS-HEP-PH-01-31",
    doi = "10.1103/PhysRevD.65.056008",
    journal = "Phys. Rev. D",
    volume = "65",
    pages = "056008",
    year = "2002"
}

@article{Zhu:1997as,
    author = "Zhu, Shi-Lin and Hwang, W-Y. P. and Yang, Ze-Sen",
    title = "{The $\Sigma_c$ and $\Lambda_c$ magnetic moments from QCD spectral sum rules}",
    eprint = "hep-ph/9708411",
    archivePrefix = "arXiv",
    doi = "10.1103/PhysRevD.56.7273",
    journal = "Phys. Rev. D",
    volume = "56",
    pages = "7273--7275",
    year = "1997"
}

@article{Ozdem:2024brk,
    author = {{\"O}zdem, Ula{\c{s}}},
    title = "{Magnetic dipole moments of the singly-heavy baryons with spin-$\frac{1}{2}$ and spin-$\frac{3}{2}$}",
    eprint = "2411.09405",
    archivePrefix = "arXiv",
    primaryClass = "hep-ph",
    doi = "10.1140/epja/s10050-025-01536-2",
    journal = "Eur. Phys. J. A",
    volume = "61",
    number = "3",
    pages = "62",
    year = "2025"
}

@article{Mohan:2022sxm,
    author = "Mohan, Binesh and S., Thejus Mary and Hazra, Avijit and Dhir, Rohit",
    title = "{Screening of the quark charge and mixing effects on transition moments and M1 decay widths of baryons}",
    eprint = "2211.16418",
    archivePrefix = "arXiv",
    primaryClass = "hep-ph",
    doi = "10.1103/PhysRevD.106.113007",
    journal = "Phys. Rev. D",
    volume = "106",
    number = "11",
    pages = "113007",
    year = "2022"
}

@article{Farhadi:2023ucs,
    author = "Farhadi, Mansour and Moosavi Nejad, S. Mohammad and Armat, A.",
    title = "{Analytical Determination of Mass and Magnetic Moment of Baryons in Diquark Model}",
    doi = "10.1007/s00601-023-01854-5",
    journal = "Few Body Syst.",
    volume = "64",
    number = "3",
    pages = "75",
    year = "2023",
    note = "[Erratum: Few Body Syst. 64, 76 (2023)]"
}

@article{Julia-Diaz:2004yqv,
    author = "Julia-Diaz, B. and Riska, D. O.",
    title = "{Baryon magnetic moments in relativistic quark models}",
    eprint = "hep-ph/0401096",
    archivePrefix = "arXiv",
    doi = "10.1016/j.nuclphysa.2004.03.078",
    journal = "Nucl. Phys. A",
    volume = "739",
    pages = "69--88",
    year = "2004"
}

@article{Faessler:2006ft,
    author = "Faessler, Amand and Gutsche, Th. and Ivanov, M. A. and Korner, J. G. and Lyubovitskij, V. E. and Nicmorus, D. and Pumsa-ard, K.",
    title = "{Magnetic moments of heavy baryons in the relativistic three-quark model}",
    eprint = "hep-ph/0602193",
    archivePrefix = "arXiv",
    doi = "10.1103/PhysRevD.73.094013",
    journal = "Phys. Rev. D",
    volume = "73",
    pages = "094013",
    year = "2006"
}

@article{Scholl:2003ip,
    author = "Scholl, Stephan and Weigel, Herbert",
    title = "{Magnetic moments of baryons with a single heavy quark}",
    eprint = "hep-ph/0312282",
    archivePrefix = "arXiv",
    doi = "10.1016/j.nuclphysa.2004.01.132",
    journal = "Nucl. Phys. A",
    volume = "735",
    pages = "163--184",
    year = "2004"
}

@article{Sharma:2010vv,
    author = "Sharma, Neetika and Dahiya, Harleen and Chatley, P. K. and Gupta, Manmohan",
    title = "{Spin $\frac12^+$, spin $\frac32^+$ and transition magnetic moments of low lying and charmed baryons}",
    eprint = "1003.4338",
    archivePrefix = "arXiv",
    primaryClass = "hep-ph",
    doi = "10.1103/PhysRevD.81.073001",
    journal = "Phys. Rev. D",
    volume = "81",
    pages = "073001",
    year = "2010"
}

@article{Bernotas:2013eia,
    author = "Bernotas, Andrius and {\v{S}}imonis, Vytautas",
    title = "{Radiative M1 transitions of heavy baryons in the bag model}",
    eprint = "1302.5918",
    archivePrefix = "arXiv",
    primaryClass = "hep-ph",
    doi = "10.1103/PhysRevD.87.074016",
    journal = "Phys. Rev. D",
    volume = "87",
    number = "7",
    pages = "074016",
    year = "2013"
}

@article{Wang:2018gpl,
    author = "Wang, Guang-Juan and Meng, Lu and Li, Hao-Song and Liu, Zhan-Wei and Zhu, Shi-Lin",
    title = "{Magnetic moments of the spin-$\frac{1}{2}$ singly charmed baryons in chiral perturbation theory}",
    eprint = "1803.00229",
    archivePrefix = "arXiv",
    primaryClass = "hep-ph",
    doi = "10.1103/PhysRevD.98.054026",
    journal = "Phys. Rev. D",
    volume = "98",
    number = "5",
    pages = "054026",
    year = "2018"
}

@article{Shi:2018rhk,
    author = "Shi, Rui-Xiang and Xiao, Yang and Geng, Li-Sheng",
    title = "{Magnetic moments of the spin-1/2 singly charmed baryons in covariant baryon chiral perturbation theory}",
    eprint = "1812.07833",
    archivePrefix = "arXiv",
    primaryClass = "hep-ph",
    doi = "10.1103/PhysRevD.100.054019",
    journal = "Phys. Rev. D",
    volume = "100",
    number = "5",
    pages = "054019",
    year = "2019"
}

@article{Aliev:2015axa,
    author = "Aliev, T. M. and Barakat, T. and Savci, M.",
    title = "{Magnetic moments of heavy $J^P = {\frac12}^+$ baryons in light cone QCD sum rules}",
    eprint = "1502.06233",
    archivePrefix = "arXiv",
    primaryClass = "hep-ph",
    reportNumber = "METU-PHYS-HEP-04-15",
    doi = "10.1103/PhysRevD.91.116008",
    journal = "Phys. Rev. D",
    volume = "91",
    number = "11",
    pages = "116008",
    year = "2015"
}

@article{Can:2021ehb,
    author = "Can, Kadir Utku",
    title = "{Lattice QCD study of the elastic and transition form factors of charmed baryons}",
    eprint = "2107.13159",
    archivePrefix = "arXiv",
    primaryClass = "hep-lat",
    doi = "10.1142/S0217751X21300131",
    journal = "Int. J. Mod. Phys. A",
    volume = "36",
    number = "23",
    pages = "2130013",
    year = "2021"
}

@article{Patel:2025txc,
    author = "Patel, Kinjal and Thakkar, Kaushal",
    title = "{Electromagnetic and weak decay of singly Heavy Baryons (Qqq)}",
    eprint = "2505.10132",
    archivePrefix = "arXiv",
    primaryClass = "hep-ph",
    doi = "10.1140/epjp/s13360-025-06398-7",
    journal = "Eur. Phys. J. Plus",
    volume = "140",
    pages = "452",
    year = "2025"
}

@article{Gandhi:2018lez,
    author = "Gandhi, Keval and Shah, Zalak and Rai, Ajay Kumar",
    title = "{Decay properties of singly charmed baryons}",
    eprint = "1811.00251",
    archivePrefix = "arXiv",
    primaryClass = "hep-ph",
    doi = "10.1140/epjp/i2018-12318-1",
    journal = "Eur. Phys. J. Plus",
    volume = "133",
    number = "12",
    pages = "512",
    year = "2018"
}

@article{Can:2013tna,
    author = "Can, K. U. and Erkol, G. and Isildak, B. and Oka, M. and Takahashi, T. T.",
    title = "{Electromagnetic structure of charmed baryons in Lattice QCD}",
    eprint = "1310.5915",
    archivePrefix = "arXiv",
    primaryClass = "hep-lat",
    doi = "10.1007/JHEP05(2014)125",
    journal = "JHEP",
    volume = "05",
    pages = "125",
    year = "2014"
}

@article{Barik:1983ics,
    author = "Barik, N. and Das, M.",
    title = "{Magnetic moments of confined quarks and baryons in an independent-quark model based on Dirac equation with power-law potential}",
    doi = "10.1103/PhysRevD.28.2823",
    journal = "Phys. Rev. D",
    volume = "28",
    pages = "2823--2829",
    year = "1983"
}

@article{Korner:1994nh,
    author = "Korner, J. G. and Kramer, M. and Pirjol, D.",
    title = "{Heavy baryons}",
    eprint = "hep-ph/9406359",
    archivePrefix = "arXiv",
    reportNumber = "DESY-94-095, MZ-THEP-94-08",
    doi = "10.1016/0146-6410(94)90053-1",
    journal = "Prog. Part. Nucl. Phys.",
    volume = "33",
    pages = "787--868",
    year = "1994"
}

@article{Kim:2018nqf,
    author = "Kim, June-Young and Kim, Hyun-Chul",
    title = "{Electromagnetic form factors of singly heavy baryons in the self-consistent SU(3) chiral quark-soliton model}",
    eprint = "1803.04069",
    archivePrefix = "arXiv",
    primaryClass = "hep-ph",
    reportNumber = "INHA-NTG-03/2018, INHA-NTG-03-2018",
    doi = "10.1103/PhysRevD.97.114009",
    journal = "Phys. Rev. D",
    volume = "97",
    number = "11",
    pages = "114009",
    year = "2018"
}

@article{Yang:2018uoj,
    author = "Yang, Ghil-Seok and Kim, Hyun-Chul",
    title = "{Magnetic moments of the lowest-lying singly heavy baryons}",
    eprint = "1802.05416",
    archivePrefix = "arXiv",
    primaryClass = "hep-ph",
    reportNumber = "INHA-NTG-02-2018",
    doi = "10.1016/j.physletb.2018.04.042",
    journal = "Phys. Lett. B",
    volume = "781",
    pages = "601--606",
    year = "2018"
}

@article{Oh:1991ws,
    author = "Oh, Yong-seok and Min, Dong-Pil and Rho, Mannque and Scoccola, Norberto N.",
    title = "{Massive quark baryons as skyrmions: Magnetic moments}",
    reportNumber = "SNUTP-91-37",
    doi = "10.1016/0375-9474(91)90458-I",
    journal = "Nucl. Phys. A",
    volume = "534",
    pages = "493--512",
    year = "1991"
}

@article{Liu:2018tqe,
    author = "Liu, Liang-Liang and Wang, Chao and Guo, Xin-Heng",
    title = "{Electromagnetic form factors of $\Lambda_c$ in the space-like momentum region}",
    eprint = "1801.08417",
    archivePrefix = "arXiv",
    primaryClass = "hep-ph",
    doi = "10.1088/1674-1137/42/10/103106",
    journal = "Chin. Phys. C",
    volume = "42",
    number = "10",
    pages = "103106",
    year = "2018"
}

@article{Franklin:1981rc,
    author = "Franklin, J. and Lichtenberg, D. B. and Namgung, W. and Carydas, D.",
    title = "{Wave Function Mixing of Flavor Degenerate Baryons}",
    reportNumber = "PRINT-81-0349 (TEMPLE)",
    doi = "10.1103/PhysRevD.24.2910",
    journal = "Phys. Rev. D",
    volume = "24",
    pages = "2910",
    year = "1981"
}

@article{Kim:2025bms,
    author = "Kim, Hyun-Chul",
    title = "{Electromagnetic and axial-vector structure of singly heavy baryons}",
    eprint = "2503.23005",
    archivePrefix = "arXiv",
    primaryClass = "hep-ph",
    doi = "10.22323/1.483.0108",
    journal = "PoS",
    volume = "QCHSC24",
    pages = "108",
    year = "2025"
}

@article{Kim:2019wbg,
    author = "Kim, June-Young and Kim, Hyun-Chul",
    title = "{Pion mass dependence of the electromagnetic form factors of singly heavy baryons}",
    eprint = "1912.01437",
    archivePrefix = "arXiv",
    primaryClass = "hep-ph",
    reportNumber = "NTG-INHA-12/2019",
    doi = "10.1093/ptep/ptab071",
    journal = "PTEP",
    volume = "2021",
    number = "6",
    pages = "063D03",
    year = "2021"
}

@article{Cloet:2014rja,
    author = {Clo{\"e}t, Ian C. and Bentz, Wolfgang and Thomas, Anthony W.},
    title = "{Role of diquark correlations and the pion cloud in nucleon elastic form factors}",
    eprint = "1405.5542",
    archivePrefix = "arXiv",
    primaryClass = "nucl-th",
    doi = "10.1103/PhysRevC.90.045202",
    journal = "Phys. Rev. C",
    volume = "90",
    pages = "045202",
    year = "2014"
}

@article{Fu:2022rkn,
    author = "Fu, Dongyan and Sun, Bao-Dong and Dong, Yubing",
    title = "{Electromagnetic and gravitational form factors of $\Delta$ resonance in a covariant quark-diquark approach}",
    eprint = "2201.08059",
    archivePrefix = "arXiv",
    primaryClass = "hep-ph",
    doi = "10.1103/PhysRevD.105.096002",
    journal = "Phys. Rev. D",
    volume = "105",
    number = "9",
    pages = "096002",
    year = "2022"
}

@article{Fu:2023ijy,
    author = "Fu, Dongyan and Wang, JiaQi and Dong, Yubing",
    title = "{Form factors of $\Omega^-$ in a covariant quark-diquark approach}",
    eprint = "2306.04869",
    archivePrefix = "arXiv",
    primaryClass = "hep-ph",
    doi = "10.1103/PhysRevD.108.076023",
    journal = "Phys. Rev. D",
    volume = "108",
    number = "7",
    pages = "076023",
    year = "2023"
}

@article{Wang:2023bjp,
    author = "Wang, JiaQi and Fu, Dongyan and Dong, Yubing",
    title = "{Form factors of decuplet baryons in a covariant quark-diquark approach}",
    eprint = "2311.07149",
    archivePrefix = "arXiv",
    primaryClass = "hep-ph",
    doi = "10.1140/epjc/s10052-024-12406-4",
    journal = "Eur. Phys. J. C",
    volume = "84",
    number = "1",
    pages = "79",
    year = "2024"
}

@article{Wang:2024abv,
    author = "Wang, Jiaqi and Fu, Dongyan and Dong, Yubing",
    title = "{A systematic study of nucleon form factors with the pion cloud effect}",
    eprint = "2410.14953",
    archivePrefix = "arXiv",
    primaryClass = "hep-ph",
    doi = "10.1140/epjc/s10052-025-14908-1",
    journal = "Eur. Phys. J. C",
    volume = "85",
    number = "11",
    pages = "1254",
    year = "2025"
}

@article{Fu:2025vkq,
    author = "Fu, Dongyan and Xie, Ju-Jun and Dong, Yubing",
    title = "{Electromagnetic form factors of $\Omega^-$ with the meson cloud in the spacelike and timelike regions}",
    eprint = "2505.03363",
    archivePrefix = "arXiv",
    primaryClass = "hep-ph",
    doi = "10.1103/7rrc-knyk",
    journal = "Phys. Rev. D",
    volume = "112",
    number = "3",
    pages = "036016",
    year = "2025"
}

@article{Meyer:1994cn,
    author = "Meyer, H.",
    title = "{The Nucleon as a relativistic quark-diquark bound state with an exchange potential}",
    eprint = "nucl-th/9407003",
    archivePrefix = "arXiv",
    reportNumber = "TPR-94-13",
    doi = "10.1016/0370-2693(94)91439-7",
    journal = "Phys. Lett. B",
    volume = "337",
    pages = "37--42",
    year = "1994"
}

@article{Capstick:1986ter,
    author = "Capstick, Simon and Isgur, Nathan",
    title = "{Baryons in a relativized quark model with chromodynamics}",
    doi = "10.1103/physrevd.34.2809",
    journal = "Phys. Rev. D",
    volume = "34",
    number = "9",
    pages = "2809--2835",
    year = "1986"
}

@article{Georgi:1990um,
    author = "Georgi, Howard",
    title = "{An Effective Field Theory for Heavy Quarks at Low-energies}",
    reportNumber = "HUTP-90/A007",
    doi = "10.1016/0370-2693(90)91128-X",
    journal = "Phys. Lett. B",
    volume = "240",
    pages = "447--450",
    year = "1990"
}

@article{Isgur:1989vq,
    author = "Isgur, Nathan and Wise, Mark B.",
    title = "{Weak Decays of Heavy Mesons in the Static Quark Approximation}",
    reportNumber = "UTPT-89-27, CALT-68-1585",
    doi = "10.1016/0370-2693(89)90566-2",
    journal = "Phys. Lett. B",
    volume = "232",
    pages = "113--117",
    year = "1989"
}

@article{Isgur:1991wq,
    author = "Isgur, Nathan and Wise, Mark B.",
    title = "{Spectroscopy with heavy quark symmetry}",
    reportNumber = "CEBAF-TH-91-01, WM-91-101, CALT-68-1704",
    doi = "10.1103/PhysRevLett.66.1130",
    journal = "Phys. Rev. Lett.",
    volume = "66",
    pages = "1130--1133",
    year = "1991"
}

@article{Punjabi:2015bba,
    author = "Punjabi, V. and Perdrisat, C. F. and Jones, M. K. and Brash, E. J. and Carlson, C. E.",
    title = "{The Structure of the Nucleon: Elastic Electromagnetic Form Factors}",
    eprint = "1503.01452",
    archivePrefix = "arXiv",
    primaryClass = "nucl-ex",
    reportNumber = "JLAB-PHY-15-2019",
    doi = "10.1140/epja/i2015-15079-x",
    journal = "Eur. Phys. J. A",
    volume = "51",
    pages = "79",
    year = "2015"
}

@article{Castellano:1973wh,
    author = "Castellano, M. and Di Giugno, G. and Humphrey, J. W. and Sassi Palmieri, E. and Troise, G. and Troya, U. and Vitale, S.",
    title = "{The reaction $e^+e^-\to p\bar{p}$ at a total energy of 2.1 GeV}",
    doi = "10.1007/BF02734600",
    journal = "Nuovo Cim. A",
    volume = "14",
    pages = "1--20",
    year = "1973"
}

@article{E835:1999mlt,
    author = "Ambrogiani, M. and others",
    collaboration = "E835",
    title = "{Measurements of the magnetic form-factor of the proton in the timelike region at large momentum transfer}",
    reportNumber = "FERMILAB-PUB-99-027-E",
    doi = "10.1103/PhysRevD.60.032002",
    journal = "Phys. Rev. D",
    volume = "60",
    pages = "032002",
    year = "1999"
}

@article{Antonelli:1998fv,
    author = "Antonelli, A. and others",
    title = "{The first measurement of the neutron electromagnetic form-factors in the timelike region}",
    doi = "10.1016/S0550-3213(98)00083-2",
    journal = "Nucl. Phys. B",
    volume = "517",
    pages = "3--35",
    year = "1998"
}

@article{Andivahis:1994rq,
    author = "Andivahis, L. and others",
    title = "{Measurements of the electric and magnetic form-factors of the proton from $Q^2=1.75$ to 8.83 (GeV/c)$^2$}",
    reportNumber = "SLAC-PUB-6462",
    doi = "10.1103/PhysRevD.50.5491",
    journal = "Phys. Rev. D",
    volume = "50",
    pages = "5491--5517",
    year = "1994"
}

@article{JeffersonLabHallA:1999epl,
    author = "Jones, M. K. and others",
    collaboration = "Jefferson Lab Hall A",
    title = "{$G_{Ep}/G_{Mp}$ ratio by
polarization transfer in $\vec ep\to e\vec p$}",
    eprint = "nucl-ex/9910005",
    archivePrefix = "arXiv",
    reportNumber = "JLAB-PHY-00-71",
    doi = "10.1103/PhysRevLett.84.1398",
    journal = "Phys. Rev. Lett.",
    volume = "84",
    pages = "1398--1402",
    year = "2000"
}

@article{JeffersonLabHallA:2001qqe,
    author = "Gayou, O. and others",
    collaboration = "Jefferson Lab Hall A",
    title = "{Measurement of $G_{Ep}/G_{Mp}$ in $\vec ep\to e\vec p$ to $Q^2=5.6$ GeV$^2$}",
    eprint = "nucl-ex/0111010",
    archivePrefix = "arXiv",
    reportNumber = "JLAB-PHY-01-63",
    doi = "10.1103/PhysRevLett.88.092301",
    journal = "Phys. Rev. Lett.",
    volume = "88",
    pages = "092301",
    year = "2002"
}

@article{Denig:2012by,
    author = "Denig, Achim and Salme, Giovanni",
    title = "{Nucleon Electromagnetic Form Factors in the Timelike Region}",
    eprint = "1210.4689",
    archivePrefix = "arXiv",
    primaryClass = "hep-ex",
    doi = "10.1016/j.ppnp.2012.09.005",
    journal = "Prog. Part. Nucl. Phys.",
    volume = "68",
    pages = "113--157",
    year = "2013"
}

@article{Lin:2021umz,
    author = "Lin, Yong-Hui and Hammer, Hans-Werner and Mei{\ss}ner, Ulf-G.",
    title = "{Dispersion-theoretical analysis of the electromagnetic form factors of the nucleon: Past, present and future}",
    eprint = "2106.06357",
    archivePrefix = "arXiv",
    primaryClass = "hep-ph",
    doi = "10.1140/epja/s10050-021-00562-0",
    journal = "Eur. Phys. J. A",
    volume = "57",
    number = "8",
    pages = "255",
    year = "2021"
}

@article{Simonis:2018rld,
    author = "Simonis, Vytautas",
    title = "{Improved predictions for magnetic moments and M1 decay widths of heavy hadrons}",
    eprint = "1803.01809",
    archivePrefix = "arXiv",
    primaryClass = "hep-ph",
    journal = "",
    month = "3",
    year = "2018"
}

@article{Hohler:1974eq,
    author = "Hohler, G. and Pietarinen, E.",
    title = "{Electromagnetic Radii of Nucleon and Pion}",
    reportNumber = "TKP 20/74",
    doi = "10.1016/0370-2693(75)90220-8",
    journal = "Phys. Lett. B",
    volume = "53",
    pages = "471--475",
    year = "1975"
}

@article{Maris:2000sk,
    author = "Maris, Pieter and Tandy, Peter C.",
    title = "{The $\pi$, $K^+$, and $K^0$ electromagnetic form-factors}",
    eprint = "nucl-th/0005015",
    archivePrefix = "arXiv",
    reportNumber = "KSU-CNR-106-00",
    doi = "10.1103/PhysRevC.62.055204",
    journal = "Phys. Rev. C",
    volume = "62",
    pages = "055204",
    year = "2000"
}

@article{Bincer:1959tz,
    author = "Bincer, Adam M.",
    title = "{Electromagnetic structure of the nucleon}",
    doi = "10.1103/PhysRev.118.855",
    journal = "Phys. Rev.",
    volume = "118",
    pages = "855--863",
    year = "1960"
}

@article{Chen:2016iyi,
    author = "Chen, Bing and Wei, Ke-Wei and Liu, Xiang and Matsuki, Takayuki",
    title = "{Low-lying charmed and charmed-strange baryon states}",
    eprint = "1609.07967",
    archivePrefix = "arXiv",
    primaryClass = "hep-ph",
    doi = "10.1140/epjc/s10052-017-4708-x",
    journal = "Eur. Phys. J. C",
    volume = "77",
    number = "3",
    pages = "154",
    year = "2017"
}

@article{Capitani:2015sba,
    author = {Capitani, S. and Della Morte, M. and Djukanovic, D. and von Hippel, G. and Hua, J. and J{\"a}ger, B. and Knippschild, B. and Meyer, H. B. and Rae, T. D. and Wittig, H.},
    title = "{Nucleon electromagnetic form factors in two-flavor QCD}",
    eprint = "1504.04628",
    archivePrefix = "arXiv",
    primaryClass = "hep-lat",
    reportNumber = "MITP-15-026, HIM-2015-01, CP3-ORIGINS-2015-012, DIAS-2015-12",
    doi = "10.1103/PhysRevD.92.054511",
    journal = "Phys. Rev. D",
    volume = "92",
    number = "5",
    pages = "054511",
    year = "2015"
}

@article{Abdel-Rehim:2015jna,
    author = "Abdel-Rehim, Abdou and Alexandrou, Constantia and Constantinou, Martha and Hadjiyiannakou, Kyriakos and Jansen, Karl and Koutsou, Giannis",
    title = "{Nucleon electromagnetic form factors from twisted mass lattice QCD}",
    eprint = "1501.01480",
    archivePrefix = "arXiv",
    primaryClass = "hep-lat",
    doi = "10.22323/1.214.0148",
    journal = "PoS",
    volume = "LATTICE2014",
    pages = "148",
    year = "2015"
}

@article{Djukanovic:2015hnh,
    author = "Djukanovic, Dalibor and Harris, Tim and von Hippel, Georg and Junnarkar, Parikshit and Meyer, Harvey B. and Wittig, Hartmut",
    title = "{Nucleon electromagnetic form factors and axial charge from CLS $N_\mathrm{f}=2+1$ ensembles}",
    eprint = "1511.07481",
    archivePrefix = "arXiv",
    primaryClass = "hep-lat",
    doi = "10.22323/1.251.0137",
    journal = "PoS",
    volume = "LATTICE2015",
    pages = "137",
    year = "2016"
}

@article{Alexandrou:2017ypw,
    author = "Alexandrou, Constantia and Constantinou, Martha and Hadjiyiannakou, Kyriakos and Jansen, Karl and Kallidonis, Christos and Koutsou, Giannis and Vaquero Aviles-Casco, Alejandro",
    title = "{Nucleon electromagnetic form factors using lattice simulations at the physical point}",
    eprint = "1706.00469",
    archivePrefix = "arXiv",
    primaryClass = "hep-lat",
    reportNumber = "DESY-17-085",
    doi = "10.1103/PhysRevD.96.034503",
    journal = "Phys. Rev. D",
    volume = "96",
    number = "3",
    pages = "034503",
    year = "2017"
}

@article{ParticleDataGroup:2024cfk,
    author = "Navas, S. and others",
    collaboration = "Particle Data Group",
    title = "{Review of particle physics}",
    doi = "10.1103/PhysRevD.110.030001",
    journal = "Phys. Rev. D",
    volume = "110",
    number = "3",
    pages = "030001",
    year = "2024"
}

@article{Pacetti:2014jai,
    author = "Pacetti, S. and Baldini Ferroli, R. and Tomasi-Gustafsson, E.",
    title = "{Proton electromagnetic form factors: Basic notions, present achievements and future perspectives}",
    doi = "10.1016/j.physrep.2014.09.005",
    journal = "Phys. Rept.",
    volume = "550-551",
    pages = "1--103",
    year = "2015"
}

@article{Ramalho:2019koj,
    author = "Ramalho, G. and Pe{\~n}a, M. T. and Tsushima, K.",
    title = "{Hyperon electromagnetic timelike elastic form factors at large $q^2$}",
    eprint = "1908.04864",
    archivePrefix = "arXiv",
    primaryClass = "hep-ph",
    reportNumber = "LFTC-19-10/48",
    doi = "10.1103/PhysRevD.101.014014",
    journal = "Phys. Rev. D",
    volume = "101",
    number = "1",
    pages = "014014",
    year = "2020"
}

@article{Korner:1976hv,
    author = "Korner, J. G. and Kuroda, M.",
    title = "{$e^+e^-$ Annihilation Into Baryon-anti-Baryon Pairs}",
    reportNumber = "DESY-76-34",
    doi = "10.1103/PhysRevD.16.2165",
    journal = "Phys. Rev. D",
    volume = "16",
    pages = "2165",
    year = "1977"
}

@article{BaBar:2005pon,
    author = "Aubert, Bernard and others",
    collaboration = "BaBar",
    title = "{A Study of $e^{+} e^{-} \to p \bar{p}$ using initial state radiation with BABAR}",
    eprint = "hep-ex/0512023",
    archivePrefix = "arXiv",
    reportNumber = "SLAC-PUB-11587, BABAR-PUB-05-050",
    doi = "10.1103/PhysRevD.73.012005",
    journal = "Phys. Rev. D",
    volume = "73",
    pages = "012005",
    year = "2006"
}

@article{Wang:2026nlr,
    author = "Wang, Jiaqi and Fu, Dongyan and Dong, Yubing",
    title = "{Form factors of $\Delta (1232)$ and the electromagnetic $N-\Delta $ transition}",
    eprint = "2602.19480",
    archivePrefix = "arXiv",
    primaryClass = "hep-ph",
    doi = "10.1140/epjc/s10052-026-15889-5",
    journal = "Eur. Phys. J. C",
    volume = "86",
    number = "6",
    pages = "688",
    year = "2026"
}

@article{Fu:2025upc,
    author = "Fu, Dongyan and Dong, Yubing and Kumano, S.",
    title = "{Transversity generalized parton distributions of the {\ensuremath{\Delta}} with the diquark spectator model}",
    eprint = "2508.15245",
    archivePrefix = "arXiv",
    primaryClass = "hep-ph",
    doi = "10.1103/d17w-jkr8",
    journal = "Phys. Rev. D",
    volume = "112",
    number = "9",
    pages = "096027",
    year = "2025"
}

@article{Sun:2017gtz,
    author = "Sun, Bao-Dong and Dong, Yu-Bing",
    title = "{$\rho$ meson unpolarized generalized parton distributions with a light-front constituent quark model}",
    eprint = "1707.03972",
    archivePrefix = "arXiv",
    primaryClass = "hep-ph",
    doi = "10.1103/PhysRevD.96.036019",
    journal = "Phys. Rev. D",
    volume = "96",
    number = "3",
    pages = "036019",
    year = "2017"
}

@article{Broniowski:2008hx,
    author = "Broniowski, Wojciech and Ruiz Arriola, Enrique",
    title = "{Gravitational and higher-order form factors of the pion in chiral quark models}",
    eprint = "0809.1744",
    archivePrefix = "arXiv",
    primaryClass = "hep-ph",
    doi = "10.1103/PhysRevD.78.094011",
    journal = "Phys. Rev. D",
    volume = "78",
    pages = "094011",
    year = "2008"
}

@article{Davidson:1994uv,
    author = "Davidson, R. M. and Ruiz Arriola, E.",
    title = "{Structure functions of pseudoscalar mesons in the SU(3) NJL model}",
    reportNumber = "NIKHEF-94-P14",
    doi = "10.1016/0370-2693(95)00091-X",
    journal = "Phys. Lett. B",
    volume = "348",
    pages = "163--169",
    year = "1995"
}

@article{Barabanov:2020jvn,
    author = "Barabanov, M. Yu. and others",
    title = "{Diquark correlations in hadron physics: Origin, impact and evidence}",
    eprint = "2008.07630",
    archivePrefix = "arXiv",
    primaryClass = "hep-ph",
    reportNumber = "NJU-INP 024/20, JLAB-PHY-21-3316",
    doi = "10.1016/j.ppnp.2020.103835",
    journal = "Prog. Part. Nucl. Phys.",
    volume = "116",
    pages = "103835",
    year = "2021"
}

@article{Santopinto:2004hw,
    author = "Santopinto, E.",
    title = "{An Interacting quark-diquark model of baryons}",
    eprint = "hep-ph/0412319",
    archivePrefix = "arXiv",
    doi = "10.1103/PhysRevC.72.022201",
    journal = "Phys. Rev. C",
    volume = "72",
    pages = "022201",
    year = "2005"
}

@article{Galata:2012xt,
    author = "Galata, Giuseppe and Santopinto, Elena",
    title = "{Hybrid quark-diquark baryon model}",
    doi = "10.1103/PhysRevC.86.045202",
    journal = "Phys. Rev. C",
    volume = "86",
    pages = "045202",
    year = "2012"
}

@article{Lichtenberg:1982jp,
    author = "Lichtenberg, D. B. and Namgung, W. and Predazzi, E. and Wills, J. G.",
    title = "{Baryon Masses in a Relativistic Quark - Diquark Model}",
    reportNumber = "IUHET 74",
    doi = "10.1103/PhysRevLett.48.1653",
    journal = "Phys. Rev. Lett.",
    volume = "48",
    pages = "1653",
    year = "1982"
}

@article{Ferretti:2011zz,
    author = "Ferretti, J. and Vassallo, A. and Santopinto, E.",
    title = "{Relativistic quark-diquark model of baryons}",
    doi = "10.1103/PhysRevC.83.065204",
    journal = "Phys. Rev. C",
    volume = "83",
    pages = "065204",
    year = "2011"
}

@article{Santopinto:2014opa,
    author = "Santopinto, E. and Ferretti, J.",
    title = {{Strange and nonstrange baryon spectra in the relativistic interacting quark-diquark model with a G{\"u}rsey and Radicati-inspired exchange interaction}},
    eprint = "1412.7571",
    archivePrefix = "arXiv",
    primaryClass = "nucl-th",
    doi = "10.1103/PhysRevC.92.025202",
    journal = "Phys. Rev. C",
    volume = "92",
    number = "2",
    pages = "025202",
    year = "2015"
}

@article{Gutierrez:2014qpa,
    author = "Gutierrez, C. and De Sanctis, M.",
    title = "{A study of a relativistic quark-diquark model for the nucleon}",
    doi = "10.1140/epja/i2014-14169-7",
    journal = "Eur. Phys. J. A",
    volume = "50",
    number = "11",
    pages = "169",
    year = "2014"
}

@article{DeSanctis:2014ria,
    author = "De Sanctis, M. and Ferretti, J. and Santopinto, E. and Vassallo, A.",
    title = "{Relativistic quark-diquark model of baryons with a spin-isospin transition interaction: Non-strange baryon spectrum and nucleon magnetic moments}",
    eprint = "1410.0590",
    archivePrefix = "arXiv",
    primaryClass = "hep-ph",
    doi = "10.1140/epja/i2016-16121-3",
    journal = "Eur. Phys. J. A",
    volume = "52",
    number = "5",
    pages = "121",
    year = "2016"
}

@article{Ebert:1996ec,
    author = "Ebert, D. and Faustov, R. N. and Galkin, V. O. and Martynenko, A. P. and Saleev, V. A.",
    title = "{Heavy baryons in the relativistic quark model}",
    eprint = "hep-ph/9607314",
    archivePrefix = "arXiv",
    reportNumber = "DESY-96-135, HUB-EP-96-25",
    doi = "10.1007/s002880050534",
    journal = "Z. Phys. C",
    volume = "76",
    pages = "111--115",
    year = "1997"
}

@article{Gershtein:2000nx,
    author = "Gershtein, S. S. and Kiselev, V. V. and Likhoded, A. K. and Onishchenko, A. I.",
    title = "{Spectroscopy of doubly heavy baryons}",
    doi = "10.1103/PhysRevD.62.054021",
    journal = "Phys. Rev. D",
    volume = "62",
    pages = "054021",
    year = "2000"
}

@article{Ebert:2002ig,
    author = "Ebert, D. and Faustov, R. N. and Galkin, V. O. and Martynenko, A. P.",
    title = "{Mass spectra of doubly heavy baryons in the relativistic quark model}",
    eprint = "hep-ph/0201217",
    archivePrefix = "arXiv",
    reportNumber = "HU-EP-02-01",
    doi = "10.1103/PhysRevD.66.014008",
    journal = "Phys. Rev. D",
    volume = "66",
    pages = "014008",
    year = "2002"
}

@article{Giannuzzi:2009gh,
    author = "Giannuzzi, Floriana",
    title = "{Doubly heavy baryons in a Salpeter model with AdS/QCD inspired potential}",
    eprint = "0902.4624",
    archivePrefix = "arXiv",
    primaryClass = "hep-ph",
    reportNumber = "BARI-TH-605-09",
    doi = "10.1103/PhysRevD.79.094002",
    journal = "Phys. Rev. D",
    volume = "79",
    pages = "094002",
    year = "2009"
}

@article{Ali:2017wsf,
    author = "Ali, Ahmed and Maiani, Luciano and Borisov, Anatoly V. and Ahmed, Ishtiaq and Jamil Aslam, M. and Parkhomenko, Alexander Ya. and Polosa, Antonio D. and Rehman, Abdur",
    title = "{A new look at the Y tetraquarks and $\Omega _c$ baryons in the diquark model}",
    eprint = "1708.04650",
    archivePrefix = "arXiv",
    primaryClass = "hep-ph",
    reportNumber = "DESY-17-104",
    doi = "10.1140/epjc/s10052-017-5501-6",
    journal = "Eur. Phys. J. C",
    volume = "78",
    number = "1",
    pages = "29",
    year = "2018"
}

@article{Santopinto:2018ljf,
    author = "Santopinto, E. and Giachino, A. and Ferretti, J. and Garc{\'\i}a-Tecocoatzi, H. and Bedolla, M. A. and Bijker, R. and Ortiz-Pacheco, E.",
    title = "{The $\varOmega _{ c}$-puzzle solved by means of quark model predictions}",
    eprint = "1811.01799",
    archivePrefix = "arXiv",
    primaryClass = "hep-ph",
    doi = "10.1140/epjc/s10052-019-7527-4",
    journal = "Eur. Phys. J. C",
    volume = "79",
    number = "12",
    pages = "1012",
    year = "2019"
}

@article{Jakob:1993th,
    author = "Jakob, R. and Kroll, P. and Schurmann, M. and Schweiger, W.",
    title = "{Octet baryon form-factors in the diquark model}",
    eprint = "hep-ph/9310227",
    archivePrefix = "arXiv",
    reportNumber = "WU-B-93-07",
    doi = "10.1007/BF01284677",
    journal = "Z. Phys. A",
    volume = "347",
    pages = "109--116",
    year = "1993"
}

@article{Keiner:1995bu,
    author = "Keiner, V.",
    title = "{Electromagnetic form-factors of the nucleon in a covariant diquark model}",
    eprint = "hep-ph/9509284",
    archivePrefix = "arXiv",
    reportNumber = "BONN-TK-95-23",
    doi = "10.1007/s002180050015",
    journal = "Z. Phys. A",
    volume = "354",
    pages = "87",
    year = "1996"
}

@article{Ma:2002ir,
    author = "Ma, Bo-Qiang and Qing, Di and Schmidt, Ivan",
    title = "{Electromagnetic form-factors of nucleons in a light cone diquark model}",
    eprint = "hep-ph/0202015",
    archivePrefix = "arXiv",
    reportNumber = "USM-TH-115",
    doi = "10.1103/PhysRevC.65.035205",
    journal = "Phys. Rev. C",
    volume = "65",
    pages = "035205",
    year = "2002"
}

@article{DeSanctis:2011zz,
    author = "De Sanctis, M. and Ferretti, J. and Santopinto, E. and Vassallo, A.",
    title = "{Electromagnetic form factors in the relativistic interacting quark-diquark model of baryons}",
    doi = "10.1103/PhysRevC.84.055201",
    journal = "Phys. Rev. C",
    volume = "84",
    pages = "055201",
    year = "2011"
}

@article{Ramalho:2008ra,
    author = "Ramalho, G. and Pena, M. T. and Gross, Franz",
    title = "{A Covariant model for the nucleon and the Delta}",
    eprint = "0803.3034",
    archivePrefix = "arXiv",
    primaryClass = "hep-ph",
    reportNumber = "JLAB-THY-08-804",
    doi = "10.1140/epja/i2008-10599-0",
    journal = "Eur. Phys. J. A",
    volume = "36",
    pages = "329--348",
    year = "2008"
}

@article{Jaffe:2004ph,
    author = "Jaffe, R. L.",
    editor = "Kunihiro, Teiji and Onogi, Tetsuya and Abuki, H. and Takahashi, Toru T.",
    title = "{Exotica}",
    eprint = "hep-ph/0409065",
    archivePrefix = "arXiv",
    reportNumber = "MIT-CTP-3538",
    doi = "10.1016/j.physrep.2004.11.005",
    journal = "Phys. Rept.",
    volume = "409",
    pages = "1--45",
    year = "2005"
}

@article{Ferretti:2019zyh,
    author = "Ferretti, J.",
    editor = "Gothe, R. and Ilieva, Y. and Mokeev, V. and Santopinto, E. and Strauch, S.",
    title = "{Effective Degrees of Freedom in Baryon and Meson Spectroscopy}",
    doi = "10.1007/s00601-019-1483-2",
    journal = "Few Body Syst.",
    volume = "60",
    number = "1",
    pages = "17",
    year = "2019"
}

@inproceedings{Lichtenberg:1996fi,
    author = "Lichtenberg, D. B. and Roncaglia, Reenato and Predazzi, Enrico",
    title = "{Diquark model of exotic mesons}",
    eprint = "hep-ph/9611428",
    archivePrefix = "arXiv",
    reportNumber = "IUHET-344, IU-NTC-96-10, DFTT-59-96",
    pages = "146--155",
    month = "11",
    year = "1996"
}

@article{Hecht:2002ej,
    author = "Hecht, M. B. and Oettel, Martin and Roberts, C. D. and Schmidt, Sebastian M. and Tandy, Peter Charles and Thomas, Anthony William",
    title = "{Nucleon mass and pion loops}",
    eprint = "nucl-th/0201084",
    archivePrefix = "arXiv",
    reportNumber = "ANL-PHY-9968-TH-2001",
    doi = "10.1103/PhysRevC.65.055204",
    journal = "Phys. Rev. C",
    volume = "65",
    pages = "055204",
    year = "2002"
}

@article{Roberts:2011cf,
    author = "Roberts, Hannes L. L. and Chang, Lei and Cloet, Ian C. and Roberts, Craig D.",
    title = "{Masses of ground and excited-state hadrons}",
    eprint = "1101.4244",
    archivePrefix = "arXiv",
    primaryClass = "nucl-th",
    doi = "10.1007/s00601-011-0225-x",
    journal = "Few Body Syst.",
    volume = "51",
    pages = "1--25",
    year = "2011"
}

@article{Burden:1996nh,
    author = "Burden, C. J. and Qian, Lu and Roberts, Craig D. and Tandy, P. C. and Thomson, Mark J.",
    title = "{Ground state spectrum of light quark mesons}",
    eprint = "nucl-th/9605027",
    archivePrefix = "arXiv",
    reportNumber = "ANL-PHY-8421-TH-96, KSUCNR-03-96",
    doi = "10.1103/PhysRevC.55.2649",
    journal = "Phys. Rev. C",
    volume = "55",
    pages = "2649--2664",
    year = "1997"
}

@article{Hess:1998sd,
    author = "Hess, M. and Karsch, F. and Laermann, E. and Wetzorke, I.",
    title = "{Diquark masses from lattice QCD}",
    eprint = "hep-lat/9804023",
    archivePrefix = "arXiv",
    reportNumber = "BI-TP-98-09",
    doi = "10.1103/PhysRevD.58.111502",
    journal = "Phys. Rev. D",
    volume = "58",
    pages = "111502",
    year = "1998"
}

@article{Maris:2002yu,
    author = "Maris, P.",
    title = "{Effective masses of diquarks}",
    eprint = "nucl-th/0204020",
    archivePrefix = "arXiv",
    doi = "10.1007/s00601-002-0111-7",
    journal = "Few Body Syst.",
    volume = "32",
    pages = "41--52",
    year = "2002"
}

@article{Ebert:2011kk,
    author = "Ebert, D. and Faustov, R. N. and Galkin, V. O.",
    title = "{Spectroscopy and Regge trajectories of heavy baryons in the relativistic quark-diquark picture}",
    eprint = "1105.0583",
    archivePrefix = "arXiv",
    primaryClass = "hep-ph",
    reportNumber = "HU-EP-11-21",
    doi = "10.1103/PhysRevD.84.014025",
    journal = "Phys. Rev. D",
    volume = "84",
    pages = "014025",
    year = "2011"
}

@article{Ebert:2005xj,
    author = "Ebert, D. and Faustov, R. N. and Galkin, V. O.",
    title = "{Masses of heavy baryons in the relativistic quark model}",
    eprint = "hep-ph/0504112",
    archivePrefix = "arXiv",
    reportNumber = "HU-EP-05-18",
    doi = "10.1103/PhysRevD.72.034026",
    journal = "Phys. Rev. D",
    volume = "72",
    pages = "034026",
    year = "2005"
}

@article{Ebert:1996ab,
    author = "Ebert, Dietmar and Feldmann, Thorsten and Kettner, Christiane and Reinhardt, Hugo",
    title = "{Heavy baryons in the quark-diquark picture}",
    eprint = "hep-ph/9601257",
    archivePrefix = "arXiv",
    reportNumber = "HUB-EP-95-28, DESY-96-010",
    doi = "10.1142/S0217751X98000482",
    journal = "Int. J. Mod. Phys. A",
    volume = "13",
    pages = "1091--1113",
    year = "1998"
}

@article{Ni:2021pce,
    author = "Ni, Ru-Hui and Li, Qi and Zhong, Xian-Hui",
    title = "{Mass spectra and strong decays of charmed and charmed-strange mesons}",
    eprint = "2110.05024",
    archivePrefix = "arXiv",
    primaryClass = "hep-ph",
    doi = "10.1103/PhysRevD.105.056006",
    journal = "Phys. Rev. D",
    volume = "105",
    number = "5",
    pages = "056006",
    year = "2022"
}

\end{document}